\documentclass[useAMS,usenatbib,letterpaper]{mn2e}
\pdfoutput=1
\usepackage{graphicx}
\usepackage[T1]{fontenc}
\usepackage{ae,aecompl}
\usepackage{color}
\usepackage{rotating}
\DeclareGraphicsExtensions{.pdf, .png, .jpg, .jpeg, .tif, .mps, .bmp, .tiff, .gif}

\usepackage[normal]{subfigure}
\usepackage[flushleft]{threeparttable}
\usepackage{graphicx}
\usepackage{amsmath}
\usepackage{amssymb}
\usepackage{multirow}
\usepackage{epstopdf}
\usepackage{float}
\usepackage{caption}
\newcommand{\ew}{\ensuremath{EW_{[OIII]}}}

\title{Orientation effects on spectral emission features of quasars}
\author[S. Bisogni et al.]{
Susanna Bisogni,$^{1,2}$\thanks{E-mail: susanna@arcetri.astro.it}
Alessandro Marconi,$^{1,2}$
and Guido Risaliti$^{1,2}$
\\
$^{1}$Dipartimento di Fisica e Astronomia, Universit\`a di Firenze, via G. Sansone 1, I-50019, Sesto Fiorentino (Firenze), Italy\\
$^{2}$INAF-Osservatorio Astrofisico di Arcetri, Largo E. Fermi 5, I-50125, Firenze\\
}

\begin{document}

\maketitle

\begin{abstract}
We present an analysis of the orientation effects in SDSS quasar composite spectra. In a previous work we have shown that the equivalent width $EW$ of the [OIII] $\lambda$5008\AA~ line is a reliable indicator of the inclination of the accretion disk.  Here, we have selected a sample of $\sim$15,000 quasars from the SDSS $7^{th}$ Data Release and divided it in sub-samples with different values of \ew. We find inclination effects both on broad and narrow quasars emission lines, among which an increasing broadening from low to high EW for the broad lines and a decreasing importance of the blue component for the narrow lines. These effects are naturally explained with a variation of source inclination from nearly face-on to edge-on, confirming the goodness of \ew as an orientation indicator. Moreover, we suggest that orientation effects could explain, at least partially, the origin of the anticorrelation between [OIII] and FeII intensities, i.e. the well known Eigenvector 1.
\end{abstract}


\section{Introduction}

The optical-UV emission of Active Galactic Nuclei (AGN) is ascribed to an accretion disk around a supermassive Black Hole (BH).
Models developed for such a structure predict that, in order to be radiatively efficient, the disk must be optically thick and geometrically thin \citep{ShakuraSunyaev1973}.
In this case the geometry of the emitting region imposes a disk continuum intensity that decreases with $\cos \theta$, $\theta$ being the angle between the disk axis and the observer line of sight, i.e. the source inclination angle. This fact can be hardly directly proven due to the difficulties in intrinsic continuum measurements. The simpler way to test the behaviour of the continuum as a function of the inclination angle is therefore a comparison between this angle-dependent continuum emission and an inclination-independent one.

The [OIII] line at $5008$\AA, emitted by the Narrow Line Region (NLR) at hundreds of parsecs from the central black hole, has isotropic characteristics, at least if compared with the emissions coming from accretion disk and BLR, and is considered a good indicator of bolometric luminosity of AGN \citep{Mulchaey94, Heckman2004}. Since line emitting regions are optically thin to line radiation, isotropy depends on their dimensions, i.e. they have to be large enough not to be significantly obscured by opaque structures, such as the accretion disk, the dusty torus and possible nuclear dust lanes. \citet{Mulchaey94} find that the [OIII] emission is isotropic in Seyfert galaxies (but see \citet{Diamond-Stanic2009} and \citet{diSeregoAlighieri97} for different results).
As a consequence, the observed \ew, i.e., the ratio between line and local (same wavelength) continuum intensities is expected to be a function of the inclination angle $\theta$.

Moreover, the [OIII] line holds a fundamental role in the context of the \emph{Eigenvector1}, the set of spectral properties able to explain most of the variance in optical spectra of quasars \citep{BorosonGreen1992}. The Eigenvector1 is usually referred to as the anticorrelation between [OIII] and FeII intensities, but other spectral properties, such as the H$\beta$ FWHM, contribute to drive the variance of quasars spectra.

In a previous work by our group (\citet{Risaliti2011}, hereafter R11) we have studied the distribution of the observed \ew\ for a flux limited sample of $\sim$7,300 quasars at redshift z$<0.8$, obtained from the SDSS DR5 quasar catalog \citep{Schneider2007}. The observed \ew\ distribution peaks at $\sim$10\AA$\;$ and appears to be dominated by the orientation effect at $EW>30$\AA, while at lower wavelengths it mainly resembles the intrinsic distribution of EW (in general, \ew\ is expected to depend on several geometrical and physical properties of the source, such as  the dimension and shape of the NLR and the spectral characteristic of the continuum emission).
The orientation signature consists in the presence, at high values, of a power-law tail ($\gamma=-3.5$) that can be  reproduced only by inclination effects.  

On the other hand, an examination  of the observed distribution of EWs of the main broad lines has revealed that in these cases the inclination effect is much weaker, if not totally absent. This result has been interpreted as a hint at a possible disk-like shape for the BLR: if the BLR and the accretion disk share the same geometrical anisotropy then the  EWs of broad lines should not show any angle dependence. Moreover, in such a case, the width of the broad lines should exhibit a trend with \ew\, specifically a broadening towards high \ew\, corresponding to \emph{edge-on} sources; indeed the dependence of broad line linewidths in terms of orientation is well known observationally \citep{WillsBrowne1986}.

 As a preliminary test, in R11 we plotted the line width of  H$\beta$ versus \ew\, and found that despite the expected large dispersion, the average widths of H$\beta$ are larger in quasars with higher \ew\, i.e. more edge-on, a fact which is difficult to explain with other scenarios.  
 
Based on these early results, the aim of this work is to search for \ew\-dependent (i.e. orientation-dependent) effects in quasars spectra.

For this purpose we have analyzed a large sample of quasars ($\sim 12000$, approximately twice as many as in R11). We divided the sample in narrow bins of \ew\, and performed a detailed spectral analysis of the staked spectra for each interval. 

The paper is organised as follows: in Section \ref{Sample} we present our new sample and we verify the presence of the relations found in R11; 
in Section \ref{sec:data_analysis} we illustrate our stacking procedure, and the spectral analysis. In Section \ref{results} we report our results on the analysis of orientation signatures in both [OIII] $\lambda 5008$\AA$\;$ and broad lines (H$\beta$, H$\alpha$ and MgII) profiles, as well as a possible explanation of the \emph{Eigenvector 1} in terms of orientation. In Section \ref{discussion} we discuss our results.

\section{Sample selection and global analysis}
\label{Sample}

Our sample of quasars has been selected from the $5^{th}$ Quasar Catalogue \citep{Schneider2010} of the SDSS $7^{th}DR$, making use of the  measured quantities listed in \citet{Shen2011}.
We first required that the redshifts of the sources were in the range $0.001<z<0.8$, in order for the [OIII] line to be into the instrumentation optimal response window. We further required a luminosity in the range of quasars ($M_{i}<-22.1$), and a signal-to-noise S/N$\geq5$. Finally we selected a range in equivalent width of the [OIII] line, between 1 and 300~\AA~ (R11), in order to avoid strong outliers probably due to incorrect measurements. The final sample includes 12,300 sources, and is a significant improvement with respect to the one presented in R11 (consisting of $\sim$7,300 sources). The increase of the source number is due  both to the use of the larger SDSS DR7 catalogue (107,850 quasars, to be compared with the $\sim$80,000 of the DR5 catalogue, used in R11), and to the inclusion of all quasars, regardless of the color selection, while in R11 we selected only objects within the ``Uniform'' subsample (Richards et al.~2006), i.e. the one selected with the standard blue excess criterion. 
This less stringent selection has the advantage of significantly increas-
\begin{center}
\includegraphics[scale=0.45]{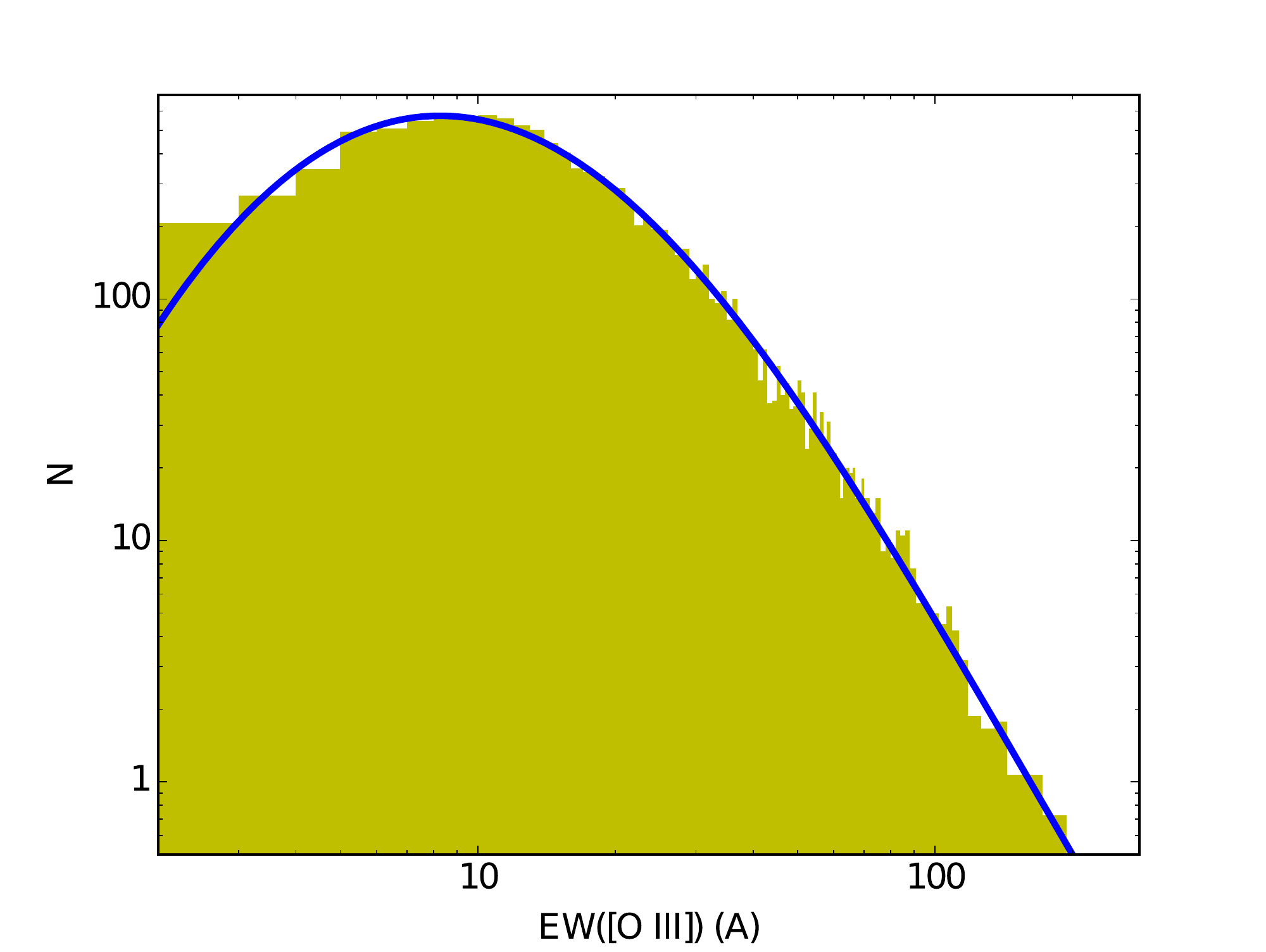}
\includegraphics[scale=0.45]{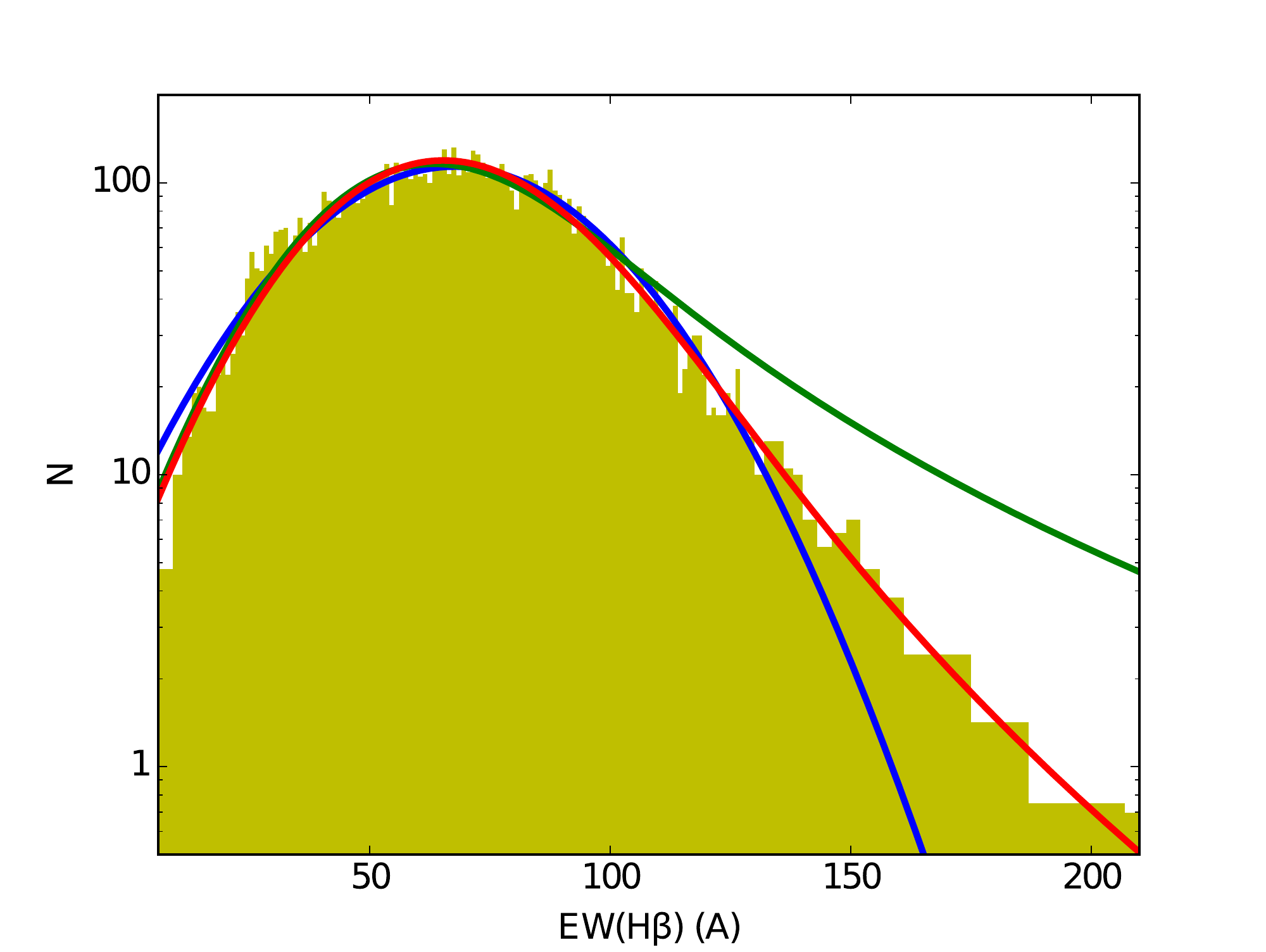}
\includegraphics[scale=0.45]{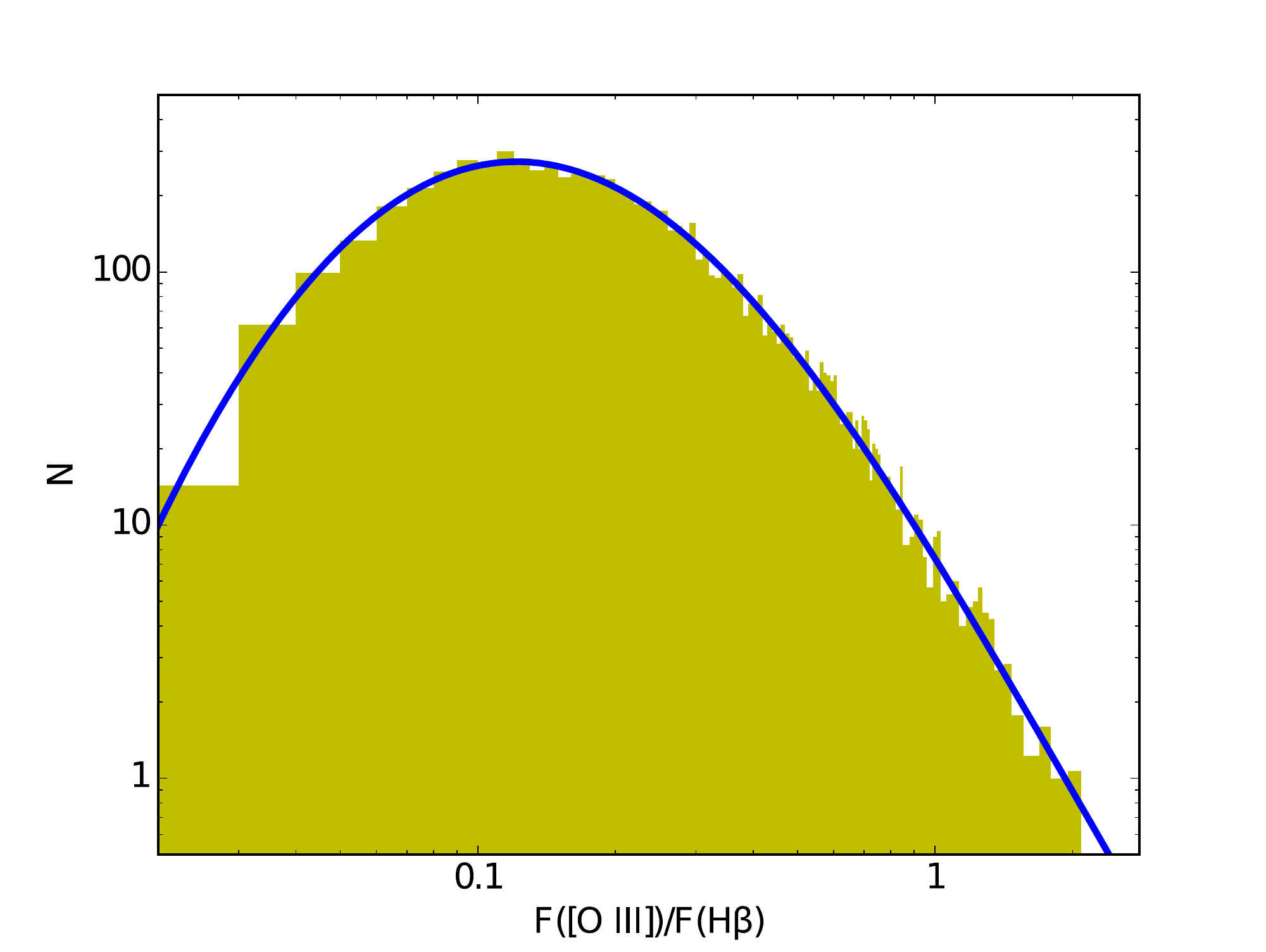}
\captionof{figure}{Observed distribution, and best fit curves, for A: \ew\; B: EW(H$\beta$); C: F([O III])/F(H$\beta$). The curves in panels B show the best fit with a high-EW tail slope $\Gamma=3.5$ (green curve), a free slope (red curve; the best fit value is $\Gamma=6.9\pm0.2$), and no tail (blue curve).}
\label{new_sample_EW_distributions}
\end{center}
ing the size of the sample, allowing a better analysis of subsamples (see next Sections), but requires a check of the global properties of this new sample.

We therefore performed a global analysis, as in R11, with some relevant modifications, as summarized below.\\
1) We analyzed the distribution of \ew, fitting the histogram in Fig. \ref{new_sample_EW_distributions}A with an integral function:
\begin{equation}
\frac{dN}{d(EW)}=\int_{0}^{EW} f(EW',L)g(EW')dL d(EW), 
\end{equation}
where L and EW' are the ``intrinsic'' (i.e. face-on) luminosity and EW of the source, f(EW',L) is a kernel which take into account for the orientation effects, and g(EW') is the intrinsic distribution of the equivalent width, which would be observed if all the objects were face-on. The derivation of f(EW',L) is presented in R11 and is not repeated here. Under the assumptions of a flux-limited sample and assuming a typical luminosity function of optically selected quasars, the integral of f(EW',L) over the luminosity produces a power law term $EW'^\Gamma$. If the distribution of inclination angles is isotropic, we obtained $\Gamma=3.5$. 
We fitted the EW distribution with a free $\Gamma$, and assuming different function for the intrinsic distribution g(EW'): a Gaussian curve, two Gaussians (as in R11), and a log-normal distribution. The latter provides the best fit to the data, with the lowest number of free parameters. The best fit values for the curve shown in Fig. \ref{new_sample_EW_distributions}, perfectly reproducing the data, are an average $\langle$log(\ew)$\rangle$ of $1.04$ and a standard deviation of $0.26$ for the intrinsic distribution and a slope of the high-EW tail $\Gamma=3.45\pm0.12$. The deviation from the intrinsic shape g(EW) starts to be relevant at EW$>30$\AA. This implies that most of the observed sources with EW$>30$\AA~ are seen nearly edge-on.\\
2) We repeated the same analysis for the equivalent width of the H$\beta$ line. The results, shown in Fig. \ref{new_sample_EW_distributions}B, are quite different from the previous case: the data are reproduced by a Gaussian distribution with average $<EW(H\beta)>$=58~\AA, standard deviation $\sigma(H\beta)$=23~\AA, convolved with a very steep high-EW tail ($\Gamma=6.9\pm0.3$). A tail with $\Gamma=3.5$ is strongly ruled out. A pure Gaussian is also statistically disfavoured, even though at a much lower significance (note the logarithmic scale in Fig.~1B). Our conclusion is that the distribution of EW(H$\beta$) suggests an almost disk-like spatial distribution of the gas emitting this line, with a small deviation with respect to the perfectly planar geometry. \\
3) As a further check of our interpretation, we analyzed the distribution of the flux ratio R=[O III]/H$\beta$. In our scheme, since the observed H$\beta$ flux has the same inclination effects as the continuum, the  ratio R should show a similar distribution as \ew. This is indeed the case, as shown in Fig. \ref{new_sample_EW_distributions}C, where the best fitting curve is obtained from the convolution of a log-normal distribution and a high-R tail with slope $\Gamma=3.20\pm0.15$. The $2\sigma$ deviation from the expected $-3.5$ value is probably an indication of the deviation of H$\beta$ emission from the pure disk-line distribution of the continuum.\\

Based on the global analysis described above and in order to investigate the spectroscopic properties of the sample as a function of disk inclination, we divided the sample in subsets of about constant \ew. We chose to create $7$ logarithmically spaced bins of \ew\, with a width of $0.3$ dex and starting form \ew $=1\AA$ (Table \ref{tab1}).
Since in the first two bins the distribution is dominated by the intrinsic dispersion rather than inclination effect (i.e we can assume that all the objects in the first $2$ bins are nearly face-on), we merged them into a single sample. We end up with six subsamples, with enough objects to ensure a proper statistical analysis.

\begin{table}
\caption{\ew\ bins. Each bin is twice the previous one in a linear scale. The widths of the bins at the edges of the distribution are larger in order to have at least $150$ objects in each bin.
$^{\star}$~The first two bins have been merged.}
\begin{center}
\begin{tabular}{c c c c c}

\hline

$\Delta$EW$_{obs}$           &    &    EW$_{obs}$        &        &    n$^{0}$ objects    \\ \relax
               [\AA]                    & $\;\;\;$   &          [\AA]             &  $\;\;\;$     &                                  \\
\hline
$1-3\;^{\star}$                                & $\;\;\;$   &   $2.0$                     &   $\;\;\;$   &       $163$    \\
$3-6\;^{\star}$                                  & $\;\;\;$ &   $4.5$                      &  $\;\;\;$   &       $932$   \\
$6-12$                            &  $\;\;\;$   &  $9.0$                       &$\;\;\;$     &       $3443$  \\ 
$12-25$                         & $\;\;\;$   &  $18.5$                       &$\;\;\;$   &       $4389$    \\
$25-50$                        & $\;\;\;$   &  $38.5$                      & $\;\;\;$   &       $2375$     \\
$50-100$                    & $\;\;\;$   &  $75.0$                       &$\;\;\;$   &       $810$        \\
$100-250$                   &$\;\;\;$  &  $175.0$                     &$\;\;\;$   &       $190$        \\
\hline
\end{tabular}
\label{tab1}
\end{center}
\end{table}

\section{Data Analysis}
\label{sec:data_analysis}

\subsection{Spectral stacking}
\label{spectra_stacking}

We produced a stacked spectrum for each of the subsamples described above.

The spectra of the individual sources within each \ew\ bin have been normalized by dividing for the flux of the line under examination (e.g. [OIII], H$\beta$, MgII), de-redshifted according to their redshift as tabulated in \citet{Shen2011} and finally rebinned to a common wavelength grid ($\lambda_{min}=2000$\AA$\;$ and $\lambda_{max}=7000$\AA$\;$ with a step of $\Delta \lambda = 0.5$\AA$\;$ to preserve the spectral resolution).
The adopted rebinning procedure is adapted to spectroscopy from the \emph{Drizzle} algorithm used for photometry of undersampled images \citep{Fruchter&Hook2002}.
The drizzle algorithm is conceived to take into account the weighted flux from each input spectrum.

The code produces the matrix of redshifted and rebinned spectra from which we can obtain the composite spectrum (\emph{stack}).

Our purpose is to obtain a precise fit of both the broad lines (H$\beta$, H$\alpha$, MgII) and the narrow line [OIII] $5008$\AA. An effect of stacking sources with different fluxes may be an alteration of the profiles of the stacked line. In order to avoid this, for each line we first normalized each spectrum to the flux of that line, and then we produced the stacked spectrum to be used on the analysis.

The average properties of each subsample are better represented by medians rather than by averages (due to the possible presence of strong outliers). However, a single median spectrum would have a too low S/N compared to the average. Based on these considerations, we obtained the stacked spectra by applying the following procedure to each spectral channel: we averaged only the fluxes between the $47^{th}$ and the $53^{rd}$ percentiles of the channel flux distribution obtained by considering all available spectra. We then
\begin{center}
\includegraphics[scale=0.34, clip=True]{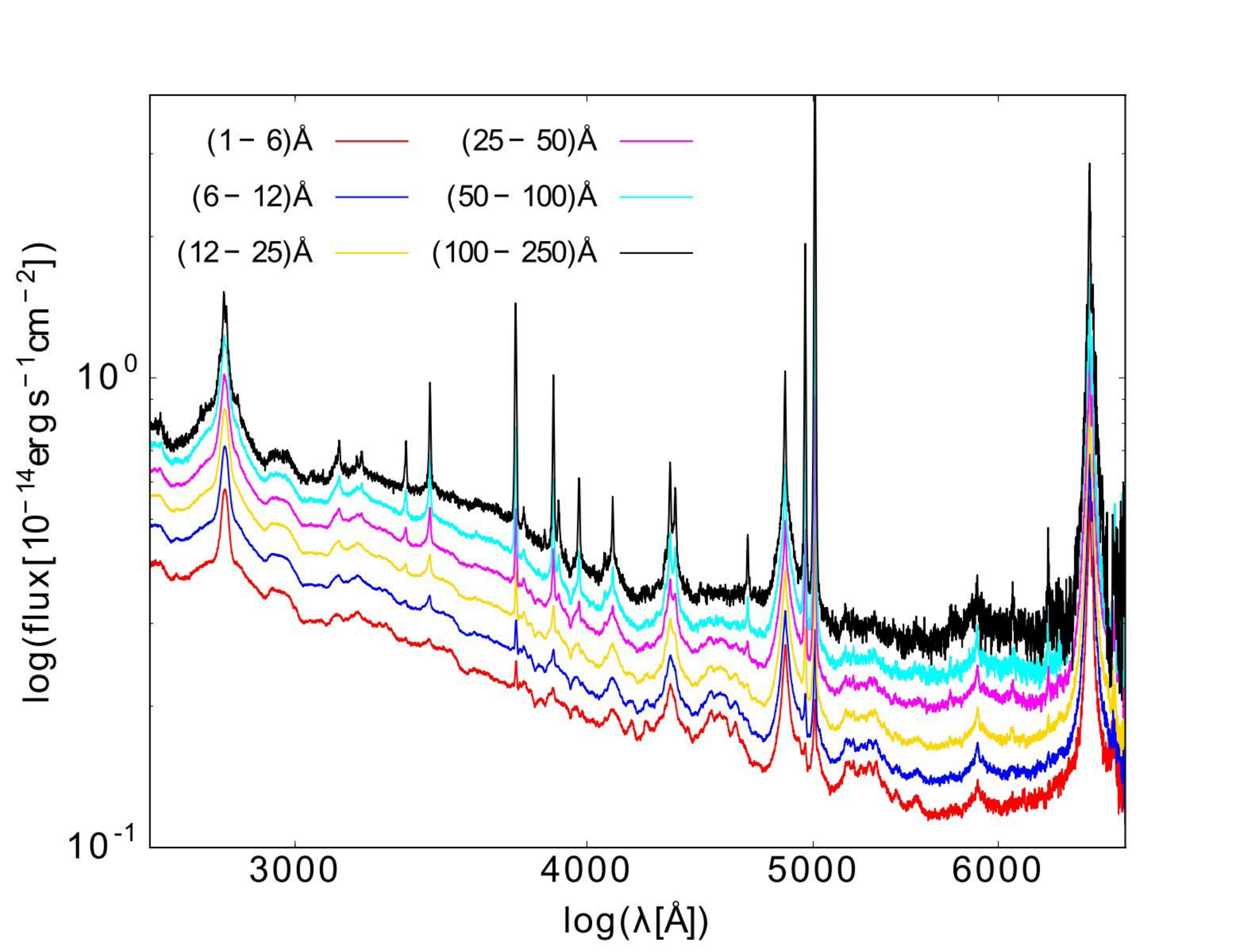}
\captionof{figure}{Stacks performed with spectra normalized to the $3000$\AA$\;$ continuum; stacks are in logarithmic scale.}
\label{stack_log3000}
\end{center}
checked that the stacks do not depend on the precise percentile levels, e.g., they are equivalent to those obtained by considering the $25^{th}$ and the $75^{th}$ percentiles. The number of spectra used to obtain the flux distribution in each spectral channel is approximately given by the number of objects presented in Tab. \ref{tab1}. Deviations from those numbers are due e.g. to flagged spectral channels in some of the spectra.
The stacked spectrum for the first \ew bin ($(1-6)$\AA) normalized to the fluxes of MgII, H$\beta$, [OIII] and H$\alpha$ respectively is reported in Tab. \ref{tab2}. The normalization of the spectra to the fluxes of a given line was adopted to preserve the average profile of that line.
An example of the final stacks is shown in Fig.\ref{stack_log3000}; the stacks are in logarithmic scale to show how the slope is always the same for all the \ew\ bins.

We have finally checked that including the not-uniform objects in the \emph{SDSS} sample (Section 2) in our selection does not affect the final results.

\begin{table}
\caption{First bin (\ew=($1-6$)\AA) stacked spectra normalized to the flux of the MgII, H$\beta$, [OIII] and H$\alpha$ line respectively. The normalization of the spectra to the fluxes of a given line was adopted to preserve the average profile of that line.}
\begin{threeparttable}
\begin{center}
\renewcommand{\arraystretch}{1.2}
\begin{tabular}{c c c c c}
\hline
                           & \multicolumn{4}{c}{$f_{\lambda}$}    \\
 $\lambda$     & norm to    &   norm to  &norm to  &  norm to    \\
                      & MgII    &      H$\beta$  &[OIII]  &  H$\alpha$    \\
\hline
   ...              &   ...         &        ...      &           ...        &     ...         \\
 $3600.0$   &  $23.4849$   &  $32.3490$ &  $349.226$ &  $10.0217$  \\
  $3600.5$  &  $23.6206$ &  $32.1183$ &$349.534$ &     $9.87436$                   \\
 $3601.0$  & $23.7292$ & $32.0483$&  $349.068$  &   $9.83273$        \\
  $3601.5$  & $23.5976$ &  $31.9613$ &  $350.720$ &  $9.78878$    \\
   $3602.0$ &  $23.1829$ &$32.0042$  & $351.764$ &   $9.86046$   \\
  $3602.5$  &  $23.2378$   & $31.9491$  &  $347.008$   &  $9.86971$     \\
   $3603.0$ & $23.5520$ &  $31.8956$  &   $346.553$  &   $9.93239$  \\   
  $3603.5$  & $23.3701$  &  $31.9325$ & $346.489$  &  $9.86804$   \\
   $3604.0$ & $23.3055$  & $32.0114$ &$346.585$ &    $9.80369$    \\
  $3604.5$  &  $23.3676$  &  $32.1024$ & $350.853$ &   $9.69592$   \\
   $3605.0$ &  $23.4790$&$31.6229$ & $350.129$ &  $9.60395$  \\
  $3605.5$  &  $23.5197$  &  $31.4075$ &  $346.206$ &    $9.66288$     \\
   $3606.0$ &  $23.1331$ & $31.5008$  &$343.109$  &  $9.79594$   \\
  $3606.5$  & $23.3483$  & $31.5023$ & $347.351$ &   $9.78815$   \\
   $3607.0$ & $23.1391$& $31.6044$ & $348.064$ &   $9.75839$   \\
  $3607.5$  &  $22.9146$ &$31.7895$  &  $344.726$  & $9.69973$   \\
   $3608.0$ &$23.0172$  &$31.6559$ &  $344.977$ & $9.65686$   \\
  $3608.5$  & $23.0082$ &$31.5133$  & $345.055$  &  $9.76075$   \\
   $3609.0$ & $23.4469$ & $31.5594$ & $347.016$ &   $9.68953$   \\
   $3609.5$ & $23.0443$  & $31.1377$& $342.511$  &  $9.66375$  \\
   $3610.0$ & $22.9955$  & $31.2424$ &  $342.452$   &   $9.59980$   \\
   ...              &   ...         &        ...      &           ...        &     ...          \\
\hline
\end{tabular}
\begin{tablenotes}
\small
\item \textbf{Notes}
\item Wavelengths are given in \AA~ and specific fluxes in $10^{-14} erg\;cm^{-2}\;s^{-1}$.
\item The complete tables for stacked spectra normalized to every line for all the \ew bins are available online.
 \end{tablenotes}
\end{center}
\end{threeparttable}
\label{tab2}
\end{table}

\subsection{Spectral fitting}
\label{sec:spec_fitting}

The IDL fitting procedure has been written making use of \emph{mpfit} \citep{Markwardt2009} which fits simultaneously continuum and spectral lines.

The template we used to fit the spectra consists of the following components:
\begin{itemize}
\item[-] a power law for the continuum
\item[-] broad emissions of FeII, obtained by convolving the individual rest frame lines inferred from the template of \citet{Veron2004} and from simulated templates for different physical conditions of the BLR obtained making use of the open source plasma simulation code Cloudy \citep{Cloudy2013}, with a Gaussian that accounts for the velocity of the emitting gas
\item[-] Broad Lines, fitted with a broken power law convolved with a Gaussian function.\\
The broken power law has the following expression
\begin{equation}
f(\lambda)\propto \begin{cases} \left(\frac{\lambda}{\lambda_{0}}\right)^{\beta} & \mbox{if}\; \lambda<\lambda_{0} \\ \left(\frac{\lambda}{\lambda_{0}}\right)^{\alpha} & \mbox{if} \; \lambda>\lambda_{0}  \end{cases}\;,
\label{eq:broaddoublepowerlawcomponent}
\end{equation}
where $\lambda_{0}$ is the central wavelength and $\alpha$ and $\beta$ are the slopes for red and blue tail respectively.
This broken power law is then convolved with a Gaussian so to avoid the presence of a cusp \citep{Nagao2006}.

\item[-] Narrow Lines, fitted with a single Gaussian, with the exception of the high ionization lines ([OIII] among them), for which an extra Gaussian is present, that takes into account the possible presence of a blue component due to outflowing gas
\end{itemize}

The fitting ranges and constraints on the parameters for all the spectral windows of interest are listed in Table \ref{tab3}. For the broad components we report the slopes of the two power laws defining the red and blue tail (respectively $\alpha$ and $\beta$), besides the line broadening, that is the $\sigma$ of the Gaussian function used to convolve the two power laws so to avoid the cuspy peaks. In the case of MgII the reported central wavelength is that of the doublet, but in the fitting procedure the two lines have been actually dealt with separately.
For narrow lines instead, fitted with a simple Gaussian function, the line broadening is the usual $\sigma$. The central wavelength reported in Tab. \ref{tab3} is the nominal laboratory vacuum wavelength for each line, used as a starting guess. The central wavelength is however left free in the fitting procedure.

The line width is expressed in $\sigma$, FWHM and \emph{Inter-Percentile Velocity} (IPV) width, defined as the difference in velocity between two reference wavelengths, $\lambda_{1}$ and $\lambda_{2}$, including a fraction $f$ of the total flux F:
\begin{equation}
\int_{\lambda_{1}}^{\lambda_{2}} f_{line}(\lambda) d\lambda = fF
\end{equation}
\citep{Whittle1985}.
The values of $\lambda_{1}$ and $\lambda_{2}$ are chosen in order to have a flux of $\frac{1-f}{2}F$ in each of the excluded tails (i.e. $\lambda<\lambda_{1}$ and $\lambda>\lambda_{2}$, respectively).
This additional quantity has been evaluated on broad lines best fit profiles to account for broadening and on [OIII] best fit profiles with the aim of defining an asymmetry coefficient (see Section \ref{results}).
 For the broad lines the width parameters are not a direct output of the fits and, for this reason, the $\sigma$, FWHM and IPVs have been computed directly from the output fitted profiles. We evaluated the errors on these quantities through a Monte Carlo simulation. In doing this we have to take into account that the (five) parameters defining a broad component are not independent. In order to produce a reasonable Monte Carlo extraction we have instead to deal with \emph{independent} parameters; so we made a change of parameters, choosing combinations that make the covariance matrix diagonal. The new independent parameters are distributed  around their central values, estimated from the old ones making use of the transformation matrix.
 We performed $1000$ extractions for each new independent parameter to collect $1000$ parameters sets from which we built a synthetic broad line. We evaluated $\sigma$ and FWHM for each realization of the line and computed the mean and the standard deviation of the $1000$ synthetic profiles, finally getting an estimate of the error on each parameter of the original profile.

\section{Results}
\label{results}

\subsection{Broad lines}
As described in Section~\ref{Sample}, the absence of the power law tail in the distribution of the EW of broad lines suggests that the BLR clouds share the same anisotropy as the accretion disk. In this case both the flux of the broad lines and that of the disk have the same dependence on the inclination angle, making the ratio between lines and continuum fluxes, i.e. the EW, no longer a function of inclination. The BLR clouds should then orbit on a disk around the SMBH and they must be optically thick to the broad emission lines. If this is the case the velocity inferred by the observer from the line widths should be only a fraction of the intrinsic one, depending on the inclination of the source with respect to the line of sight (specifically $v_{obs} = v_{int} \sin \theta$).\\
The subsequent step of our analysis has been the study of profiles as a function of \ew\ for H$\beta$, H$\alpha$ and MgII.
Fig. \ref{broad_lines_profiles} shows the broad profiles normalized to their peak values for each subsample with a constant \ew.

\begin{center}
\includegraphics[scale=1.09, clip=True]{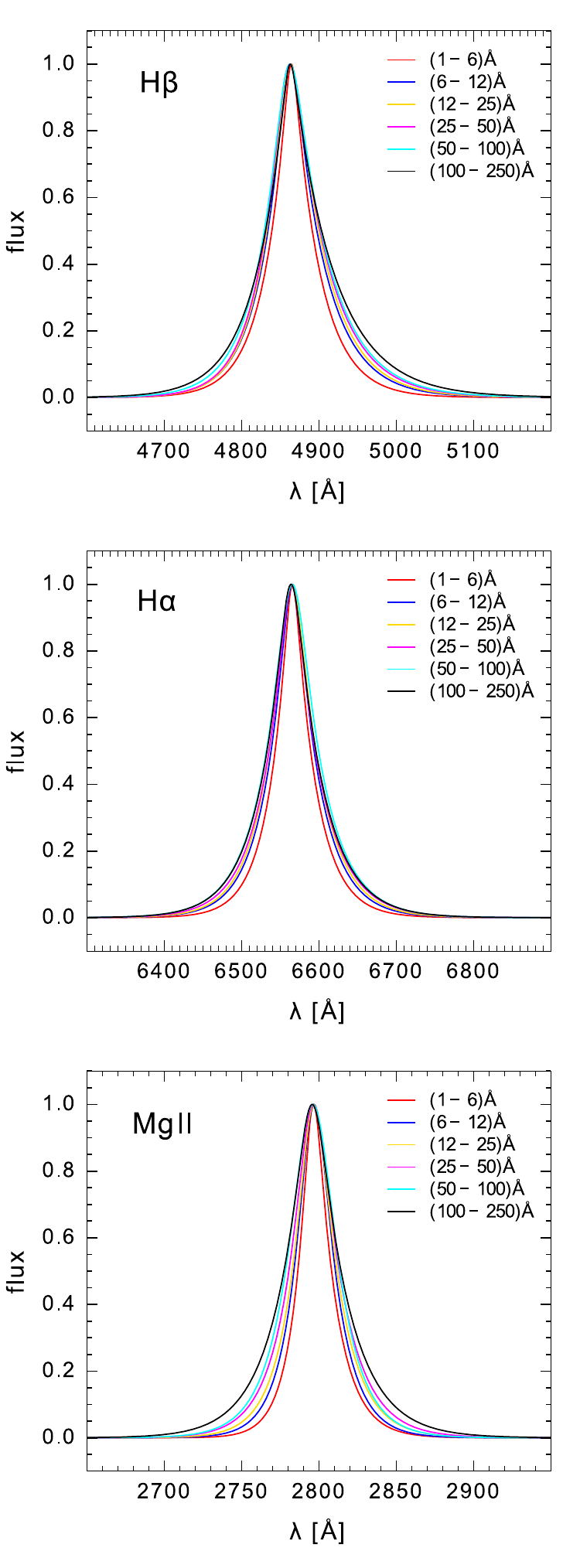}
\captionof{figure}{Broad components profiles; from top to bottom H$\beta$, H$\alpha$ and MgII. Profiles are normalized to their peak values.}
\label{broad_lines_profiles}
\end{center}

\begin{table*}
\caption{Spectral ranges and constraints on the free parameters of our template. Fitting functions: BPL is the Broken Power Law and DG indicates the use of two Gaussians to fit the line profile.}
\begin{center}
\renewcommand{\arraystretch}{1.2}
\begin{tabular}{l c c c c c c }
\hline
 line         &fitting function & fitting range&starting $\lambda_{0}$ & $\alpha$ range    &   $\beta$ range     & line broadening  range    \\
                &                                                &        (\AA)         &          (\AA)       &                           &                      &   ($km/s$)\\
\hline
H$\beta $ & BPL & $(4400-5400)$ &     $4862.68$     &  $10-500$      &$10-500$      &    $250-10000$   \\
H$\alpha$ & BPL &$(6000-7000)$ &     $6564.61$    &  $10-500$       &$10-500$       &     $100-10000$  \\
MgII           & BPL &$(2500-3100)$ &     $2798.75$      & $10-500$          &$10-500$      &  $70-10000$ \\
{[OIII]}       & DG &$(4400-5400)$&        $5008.24$     &                            &                    &      $60- $             \\
{[OII]}        & DG  &$(3650-3850)$ &     $3728.48$     &                               &                    &                   \\
\hline
\end{tabular}
\label{tab3}
\end{center}
\end{table*}

The dependence of line width with \ew\ can be interpreted, within our scenario, as due to the different source inclinations with respect to the observer line of sight in each subsample.

\begin{figure*}
\centering
\includegraphics[scale=1.0, clip=True]{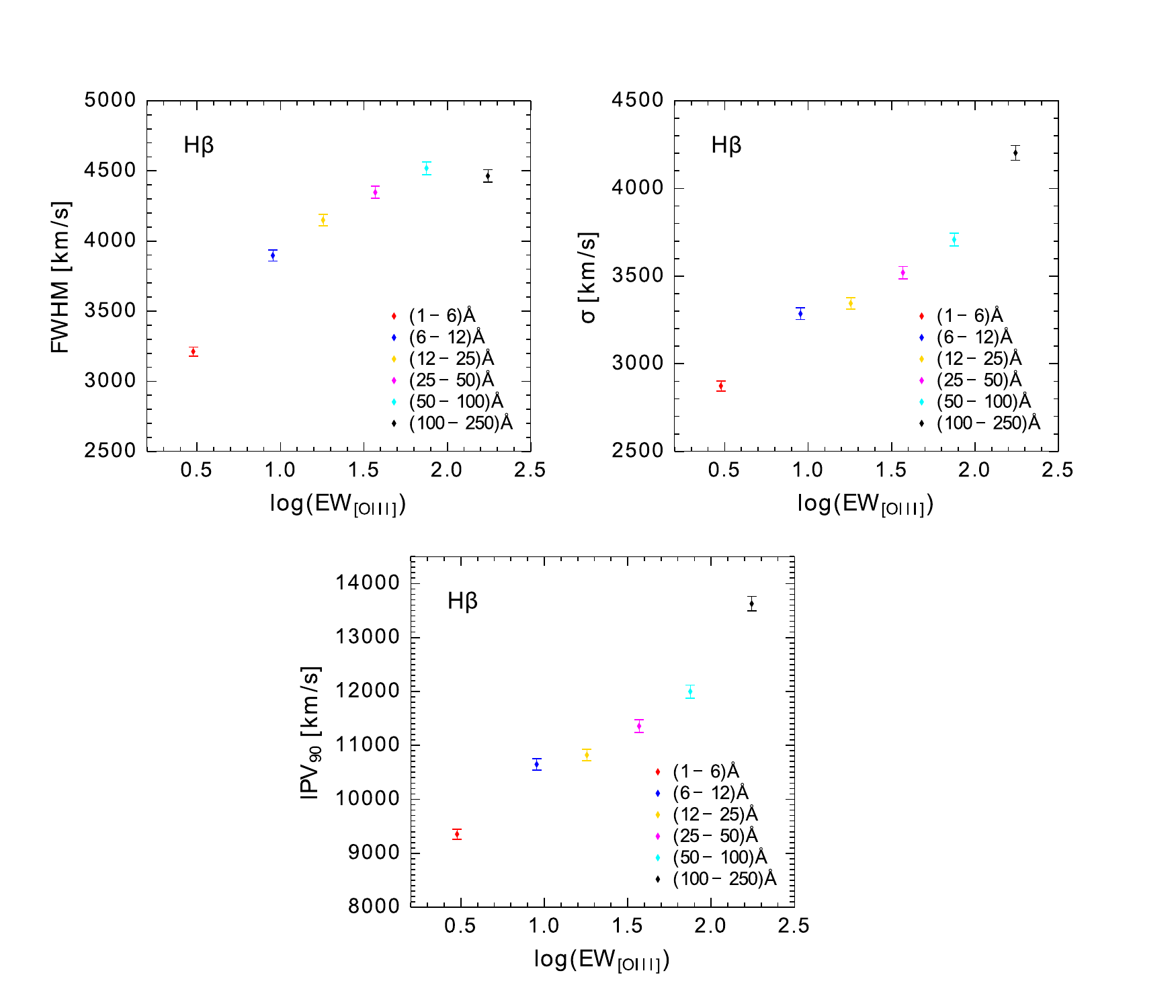}
\caption{FWHM, $\sigma$ and IPV ($90\%$) of the H$\beta$ broad component as a function of \ew.}
\label{FWHM_sigma_IPV_Hb}
\end{figure*}
\begin{figure*}
\centering
\includegraphics[scale=1.0, clip=True]{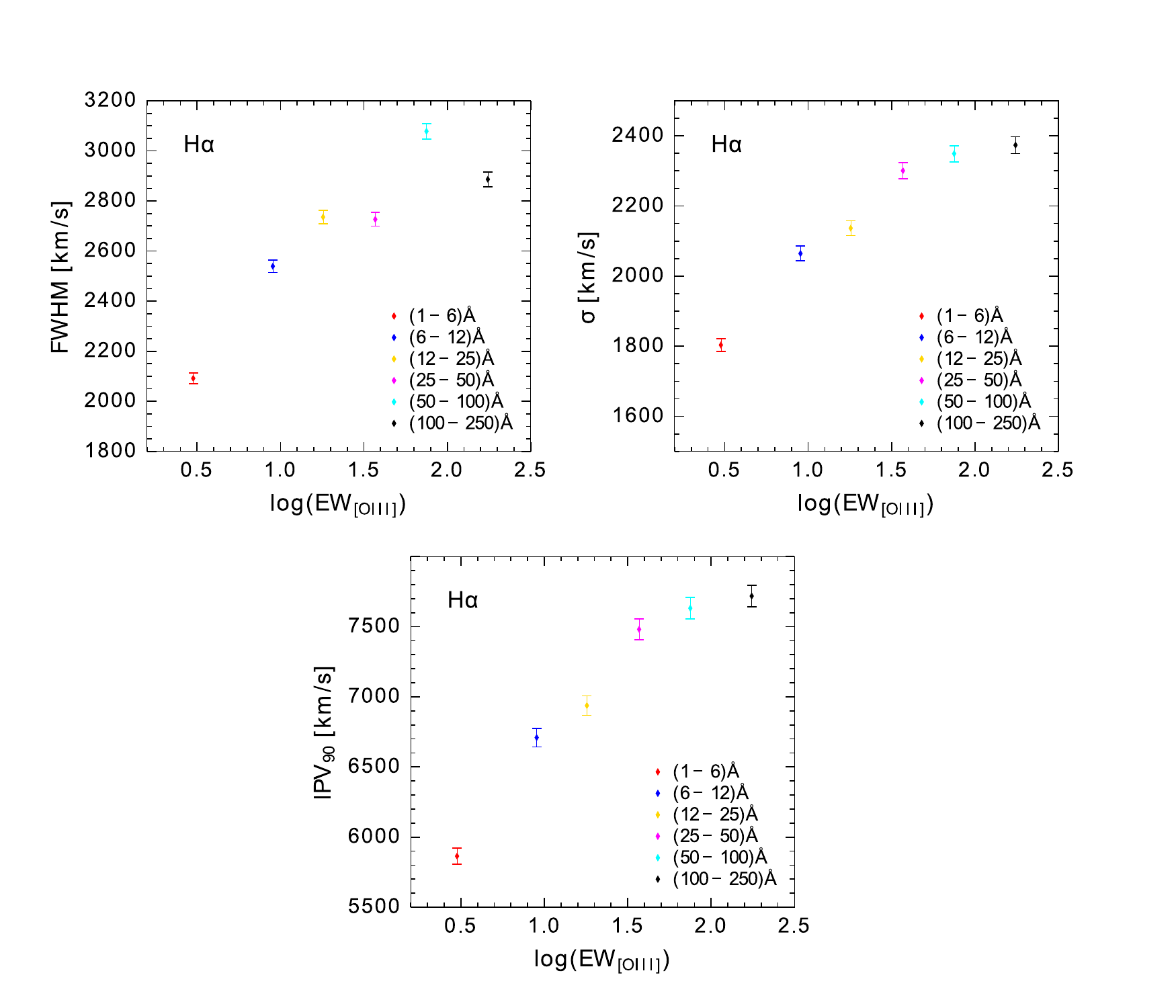}
\caption{FWHM, $\sigma$ and IPV ($90\%$) of the H$\alpha$ broad component depending on \ew.}
\label{FWHM_sigma_IPV_Ha}
\end{figure*}
\begin{figure*}
\centering
\includegraphics[scale=1.0, clip=True]{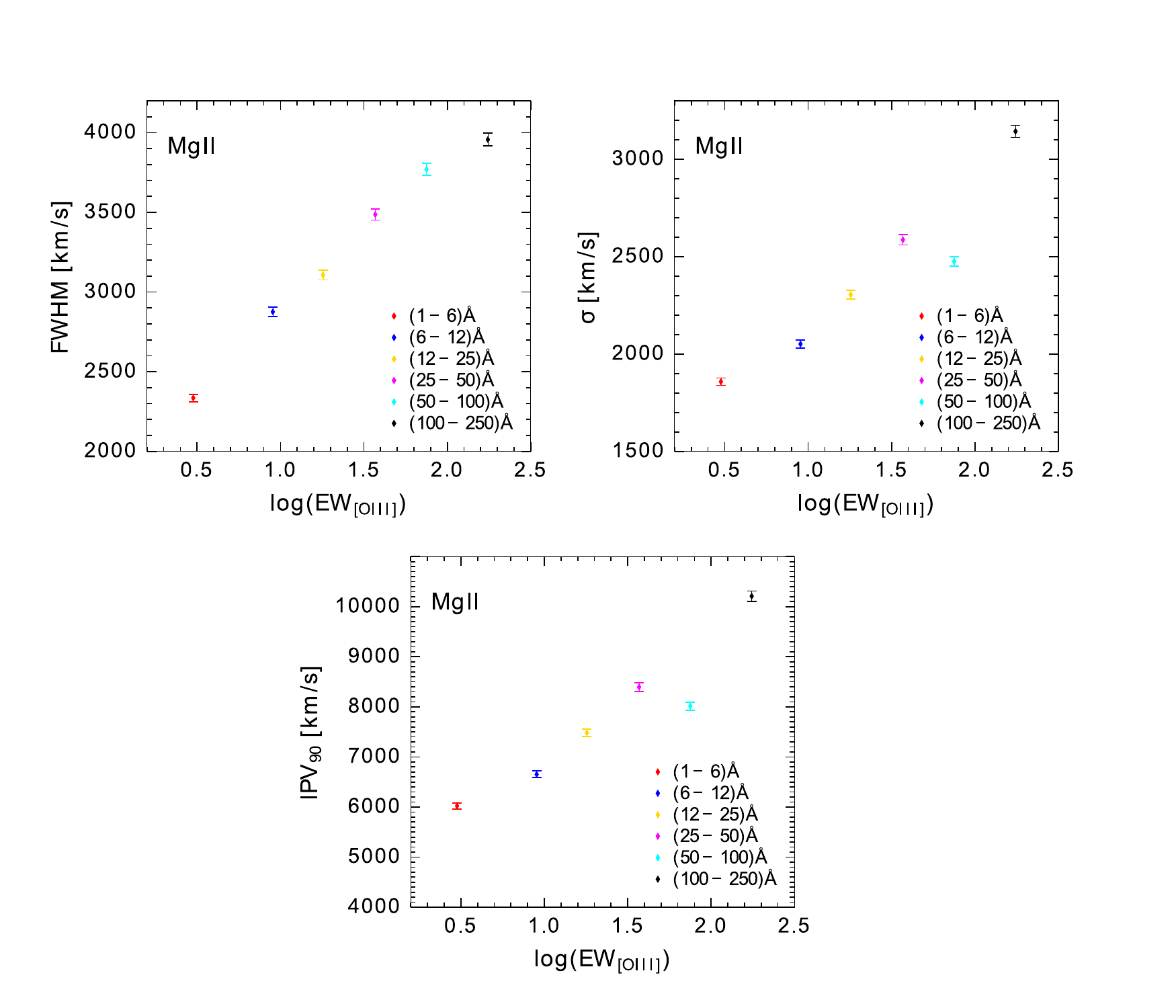}
\caption{FWHM, $\sigma$ and IPV ($90\%$) of the MgII broad component depending on \ew.}
\label{FWHM_sigma_IPV_MgII}
\end{figure*}

In Fig. \ref{FWHM_sigma_IPV_Hb}, \ref{FWHM_sigma_IPV_Ha} and \ref{FWHM_sigma_IPV_MgII} the broadening can be observed more quantitatively; these figures show the increase in FWHM, $\sigma$ and the Inter-Percentile Velocity (IPV) width (in this case we have evaluated in particular the difference between velocities corresponding to $5\%$ and $95\%$ of the integrated line flux, IPV $90\%$) for the broad components of the three emission lines as a function of \ew.
All the parameters grow rather steadily going from low (near the face-on position) to high EWs (near the edge-on position).
FWHM, $\sigma$ and IPV have been computed on the best fit line profiles. Statistical errors for these quantities have been evaluated as explained in Section \ref{sec:spec_fitting}, but due to their small values (ranging from a few $10^{-5}$ up to a few $10^{-3}$) we have conservatively assigned a $1\%$  systematic error to each of these quantities.
 Values and errors for each stack are listed in Tab. \ref{tab4}.\\

\subsection{Narrow Lines}

We then performed a detailed study of the [OIII] $\lambda 5008$\AA$\;$ line, the most prominent NLR emission feature.
Fig. \ref{OIII_profile} shows the [OIII] profile obtained for each \ew\ stack.
The blue component decreases going from low EWs (``face-on'' positions) to high EWs (edge-on positions). This behaviour is what is expected if we associate the blue component to an outflow: the preferential direction of outflows is perpendicular to the accretion disk and so the observed outflow velocity will correspond to the intrinsic one only in the face-on position, while in any other position this will be decreased by the factor $\cos \theta$.
An analysis of the shifts in the velocity of the two (main and blue) [OIII] components has also been performed: we have examined the shift of the central velocity of the two [OIII] components with respect to the [OII] doublet $\lambda \lambda 3727.092$\AA,$3729.875$\AA$\;$ velocity in each stack ($v_{[OIII]}-v_{[OII]}$), assumed to be the systemic velocity of the ``host galaxy'' of the relative stack, as first suggested by \citet{Boroson2011} (Fig. \ref{voiii-voii}). 
The shift in velocity clearly decreases going from low EWs to high EWs with a higher absolute value for the blue component with respect to the main one.
The shifts and their errors (converted to velocity shifts starting from wavelength shifts from the Gaussian curve fits) are listed in Tab. \ref{tab5}.

\subsection{FeII}
FeII emission originates from the BLR. The intensity of these lines is quite weak if compared to other broad lines; in addition, because of the presence of multiplets, they are often blended and, therefore, difficult to disentangle from each other in our spectra.
We paid special attention to the analysis of those features due to the importance they hold in the \emph{Eigenvector1}, i.e. the anticorrelation between FeII and [OIII] intensities in quasars spectra, responsible for most of the spectral variance in optical spectra of quasars \citep{BorosonGreen1992,SulenticMarziani2015}.

This anticorrelation is commonly seen in quasar spectra so its presence in our spectra stacks is not surprising.
What could be more interesting is the fact this anticorrelation has a trend with the observed \ew; moving from stacks corresponding to low \ew\ values to stacks corresponding to higher values, FeII emission becomes less and less intense, while [OIII] line behaves just the opposite.
This trend discloses a dependence of Eigenvector1, at least partially, on orientation. In low \ew\ stacks (``face-on'' objects) the BLR is face-on and so the FeII emission is seen at the peak of its intensity. 
Recently \citet{ShenHo2014} examined the relationship among the strengths of FeII and [OIII] with the H$\beta$ linewidth in the context of Eigenvector1. They ascribe the negative correlation between the H$\beta$ linewidth and the FeII strength to the BH mass or, equivalently, the Eddington Ratio.
In their scenario this property is the main driver of EV1, with the orientation only contributing to the dispersion of the linewidths at fixed FeII strength.
We find similar trends between these spectral characteristics, but the main difference in our interpretation is that we believe the orientation to be one of the drivers of the EV1 and explain the anticorrelation of the H$\beta$ linewidth and FeII strength as an orientation effect. 
We note that other physical drivers of the EV1, such as the BH mass, luminosity or Eddington ratio, may be relevant. In our analysis this effects, if present, are diluited by the average of the whole quasar population in each bin of \ew.

\begin{table*}
\caption{ $(4400-5400)$\AA~ slope continuum and broad component best fit FWHM, $\sigma$ and IPV($90\%$) of H$\beta$, H$\alpha$ and MgII for each \ew\ bin.}
\begin{center}
\renewcommand{\arraystretch}{1.2}
\begin{tabular}{ l c c c c c c c }
\hline
\ew\ (\AA)            & $[1-6]$  &  $[6-12]$ & $[12-25]$ & $[25-50]$ & $[50-100]$  & $[100-250]$ \\                  
\hline
$(4400-5400)$\AA~ slope &$-1.175\pm 0.001$& $-1.393\pm0.001$ &$-1.370\pm0.001$ &$-1.208\pm0.002$&$-1.152\pm0.003$&$-1.151\pm0.007$ \\
H$\beta$ FWHM (km/s) &$3212\pm32$ & $3897\pm38$&$4150\pm41$&$4347\pm43$&$4520\pm45$&$4464\pm45$ \\
H$\beta \; \sigma$ (km/s)&$2874\pm29$& $3285\pm33$&$3345\pm33$&$3520\pm35$&$3708\pm37$&$4203\pm42$ \\
H$\beta $ IPV($90\%$) (km/s)&$9353\pm94$& $10648\pm106$&$10821\pm108$&$11357\pm114$&$11999\pm120$&$13626\pm136$ \\
H$\alpha$ FWHM (km/s)&$2092\pm21$& $2539\pm25$&$2736\pm27$&$2727\pm27$&$3078\pm31$&$2886\pm29$ \\
H$\alpha \; \sigma$ (km/s) &$1803\pm18$ & $2064\pm21$&$2137\pm21$&$2301\pm23$&$2349\pm23$&$2374\pm24$ \\
H$\alpha$ IPV ($90\%$) (km/s) &$5864\pm59$ & $6709\pm67$&$6938\pm69$&$7481\pm75$&$7632\pm76$&$7719\pm77$ \\
MgII FWHM (km/s)   &$2335\pm23$& $2876\pm29$&$3107\pm31$&$3487\pm35$&$3770\pm38$&$3956\pm40$ \\
MgII $\sigma$ (km/s) &$1850\pm19$ & $2051\pm21$&$2305\pm23$&$2586\pm26$&$2475\pm25$&$3143\pm31$ \\
MgII IPV ($90\%$) (km/s) &$6021\pm60$ & $6652\pm67$&$7482\pm75$&$8395\pm84$&$8016\pm80$&$10209\pm102$ \\
\hline
\end{tabular}
\label{tab4}
\end{center}
\end{table*}

\begin{table*} 
\caption{Velocity shifts for main and blue [OIII] components with respect to the central velocity of [OII] (representing the systemic velocity for the host galaxy), IPV($90\%$) and asymmetry index for the total [OIII] profile for each \ew\ bin.}
\begin{center}
\renewcommand{\arraystretch}{1.3}
\begin{tabular}{ l | c c c c c c }

\hline
\ew\ (\AA)    & $[1-6]$  &  $[6-12]$ & $[12-25]$ & $[25-50]$ & $[50-100]$  & $[100-250]$ \\                         
\hline
[OIII] main shift (km/s)&$-70\pm4$& $-74\pm2$&$-63\pm1$&$-36\pm1$&$-25\pm1$&$-17\pm1$ \\ \relax
[OIII] blue shift (km/s)&$-342\pm7$& $-343\pm3$&$-260\pm2$&$-208\pm1$&$-128\pm1$&$-78\pm2$ \\
IPV($90\%$) (km/s)&$1606\pm16$& $1320\pm13$&$1722\pm17$&$1091\pm11$&$914\pm9$&$833\pm8$ \\
$A_{IPV_{05-95}}$ & $1.547\pm0.015$  & $1.686\pm0.017$  & $1.431\pm0.014$  & $1.513\pm0.015$  & $1.328\pm0.013$   & $1.173\pm0.011$   \\
\hline
\end{tabular}
\label{tab5}
\end{center}
\end{table*}

\subsection{Double-peaked broad lines}

If the H$\beta$ line is emitted by a flat distribution of clouds, we expect that double-peaked lines are more frequent in edge-on sources. In order to investigate this point, we considered the \ew\ distribution of quasars with double peaked lines in our sample. Shen et al.~(2010) provide a flag indicating ``unambiguous'', and ``possible'' double-peaked lines, based on visual inspection. Given the non-homogeneous nature of this selection a large spread in the distribution is expected; however, the relatively large number of objects (about 100 ``unambiguous'' and 500 ``possible'') allows for a comparison with the global distribution of \ew. In Fig.~\ref{double} we show the cumulative \ew\ distributions for the global sample, and the two ``unambiguous'' and ``possible'' double-peaked subsamples. It is clear that the double peaked objects are on average shifted towards higher values of \ew. We performed a Kolmogorov-Smirnov test on the distributions and we found that the probability that the shift towards high \ew\ in double peaked objects is NOT significant is below 10$^{-4}$ for both the ``unambiguous'' and the ``possible'' double-peaked quasars.

\subsection{$L_{[OIII]}$ and $L_{5100}$ distributions}

As a final point in examining the \ew\ bins we consider the luminosity distributions of [OIII] and continuum at $5100$\AA$\;$ (Fig. \ref{hist_Loiii_L5100}).
The flux limited selection is visible in the continuum luminosity distribution that is steadily peaked around $log(L_{5100}) \sim 44.6$ although it can be recognized a small shift in the central values, ascribable to the expected decrease in continuum luminosity moving from face-on to edge-on positions.
A more relevant effect of the flux limited selection is present in the [OIII] luminosity distribution: sources in edge-on position are selected only if they are intrinsically more luminous. Since [OIII] is an isotropic indicator of the intrinsic disk luminosity we observe an excess of high $L_{[OIII]}$ in object with high \ew.
The average values of $L_{[OIII]}$ and $L_{5100}$ for each bin are listed in Tab. \ref{tab6}.

\begin{table*} 
\caption{Mean values of [OIII] and continuum at $5100$\AA$\;$ luminosities distributions for each \ew\ bin.}
\begin{center}
\renewcommand{\arraystretch}{1.3}
\begin{tabular}{ l | c c c c c c }
\hline
\ew\ (\AA)    & $(1-6)$  &  $(6-12)$ & $(12-25)$ & $(25-50)$ & $(50-100)$  & $(100-250)$ \\                         
\hline
$\overline{log(L_{[OIII]})}\;\;  (erg s^{-1})$&$41.5\pm0.4$& $41.9\pm0.3$&$42.1\pm0.4$&$42.4\pm0.3$&$42.6\pm0.3$&$42.9\pm0.3$ \\ \relax
$\overline{log(L_{5100})}\;\;  (erg s^{-1})$&$44.6\pm0.4$& $44.6\pm0.3$&$44.6\pm0.3$&$44.6\pm0.3$&$44.5\pm0.3$&$44.5\pm0.3$ \\
\hline
\end{tabular}
\label{tab6}
\end{center}
\end{table*}

\section{Discussion}
\label{discussion}

Our analysis of the EW distribution of [OIII] and of the stacked spectra of quasars with different \ew\ have revealed several relevant properties of the narrow and broad line regions:
\begin{enumerate}
\item The distribution of \ew\ is well reproduced by an intrinsic log-normal distribution convolved with a high-EW power law tail. The maximum and dispersion of the intrinsic distribution are EW$_{MAX}=1$\AA~ and $\sigma = 9$\AA. The exponent of the power law tail is $\Gamma = 3.5$.
\item The distribution of EW of H$\beta$ is well represented by a Gaussian distribution with EW$_{MAX}=58$~\AA~ and $\sigma = 23$~\AA~ and a power law tail with $\Gamma \simeq 7$.
\item The [OIII]/H$\beta$ distribution shows the same high end tail as \ew.
\item The [OIII] line shows a blue tail whose intensity decreases moving from low to high \ew. The blueshift similarly decreases with EW.
\item The width of broad lines increases moving from low to high \ew.
\item  FeII emission is prominent for low \ew\ and its intensity decreases moving to high EW([OIII]).
\item Double peaked broad lines objects are more frequent for high EW([OIII]) with respect to low \ew.
\end{enumerate}

Here we discuss the physical consequences of our results.

\begin{figure}
\centering
\includegraphics[scale=0.65]{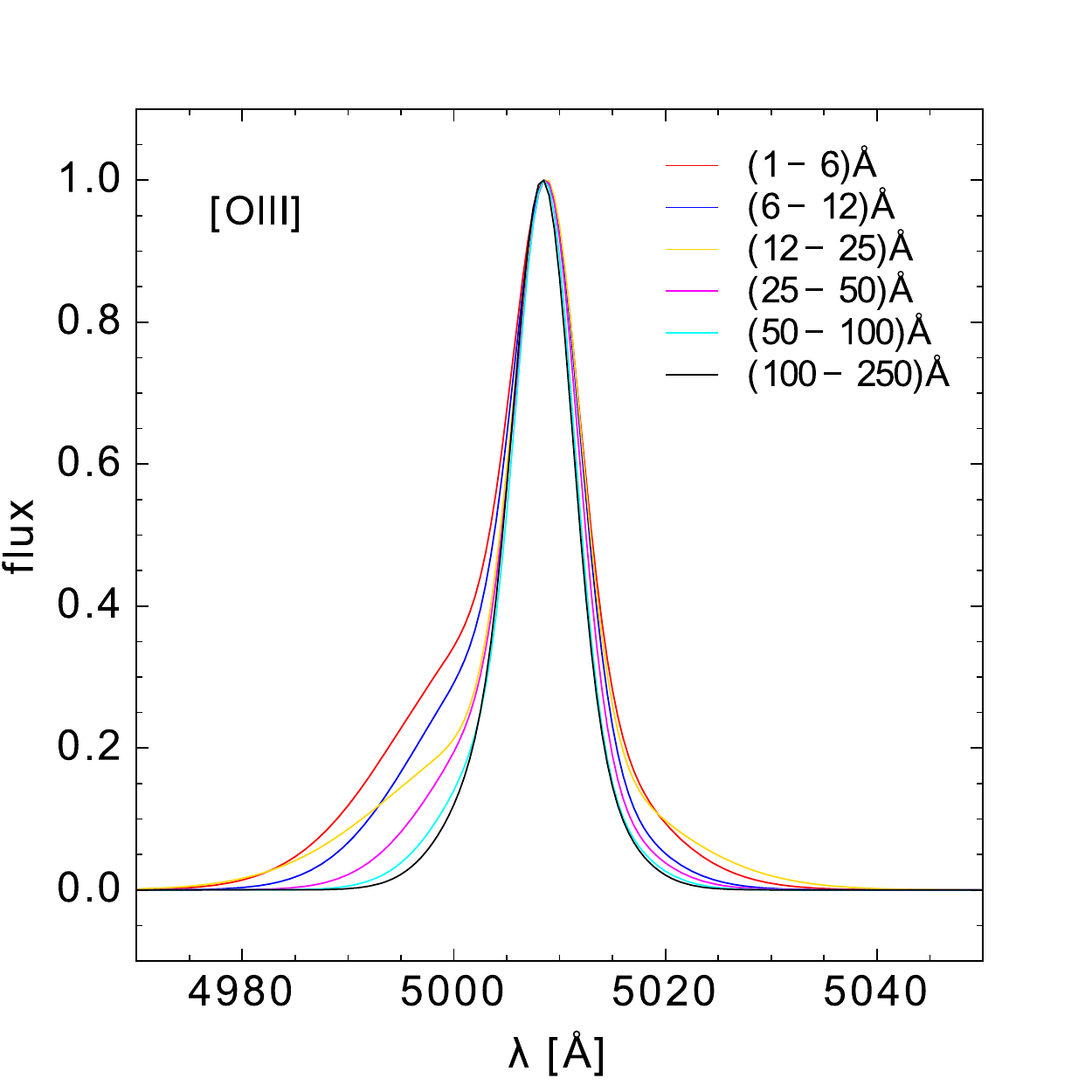}
\caption{[OIII] $\lambda 5008$\AA$\;$ profile for each \ew\ representative spectrum. Profiles are normalized to their peak values.}
\label{OIII_profile}
\end{figure}

\subsection{Distribution of \ew\ and EW([H$\beta$])}

The distribution of \ew\ is  relevant in two respects: (1) the high-EW tail $\Gamma=3.5$ fully confirms the scenario of an isotropic emission of the [O III]~$\lambda$5008~\AA~ line, with intensity proportional to the illumination from the ionizing source, and a disc-like continuum emission; (2) the ratio between the width of the intrinsic distribution and its peak value is an estimate of the precision of the [O III] luminosity as an indicator of the bolometric luminosity (which is supposed to be dominated by the disk emission). From our results, we conclude that an estimate of the bolometric/disk luminosity based on the [O III] line has an uncertainty of a factor of $\sim$2. The possible reasons for the observed intrinsic dispersion of \ew\ are variations in the covering factor of the Narrow-Line Region clouds as seen from the disc and effects of the dispersion in the optical/UV spectral energy distribution (the emission of the [O III] line is expected to be proportional to the disk emission at the line {\em ionizing} frequency of [O III], i.e. $\sim50$~eV, while the continuum is measured at the {\em emission} frequency). These effects are discussed in Risaliti et al.~(2011).
We note that in principle an increase in \ew\ could be due to dust reddening of the disc component. In this case, however, the effect of dust reddening should be seen also on continuum spectra, while we have shown in Section \ref{spectra_stacking} that our selection ensures that only blue objects are present (see also Tab. \ref{tab4}).

The distribution of EW(H$\beta$)  strongly suggests a disc-like emission of the line.
The orientation effects found in the broad emission lines require the BLR to be not only flat, but also optically thick to these lines. This is likely to be the case for the H$\beta$ line: for densities and column densities typical of BLR clouds ($n>10^{9}-10^{10}$~cm$^{-3}$ and $N_{H}>10^{23}$~cm$^{-2}$) the optical depth of the Ly$\alpha$ line is expected to be higher than $10^{4}$, and the optical depth of the Balmer lines start to be significant when $\tau$(Ly$\alpha)$ is higher than a few hundred \citep{OsterbrockFerland2006}.

Moreover, the distribution  of R=[OIII]/H$\beta$ confirms the suggestion of a disk-like shape for the H$\beta$ emitting region. If the BLR geometry resembles that of the accretion disk then this ratio is a close version of the \ew\, the difference between the two being determined by the larger height scale of the disk of the BLR, probably caused by the presence of turbulence in the gas componing this structure.

\begin{figure*}
\centering
\includegraphics[scale=0.9]{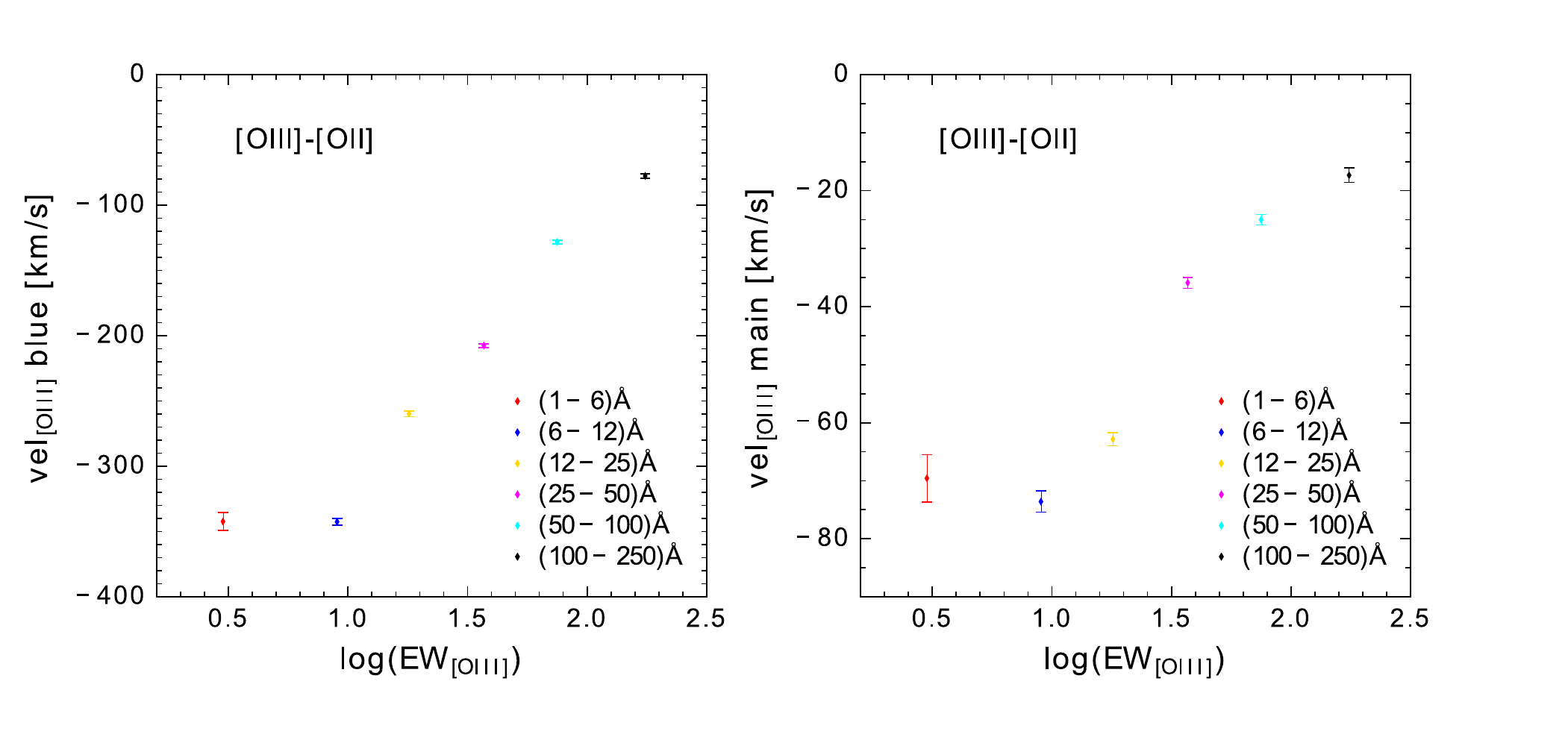}
\caption{Velocity shifts for the main (left panel) and blue (right panel) [OIII] components with respect to [OII] velocity, representing the systemic velocity for the host galaxy.}
\label{voiii-voii}
\end{figure*}

\begin{figure}
\begin{center}
\includegraphics[scale=0.4]{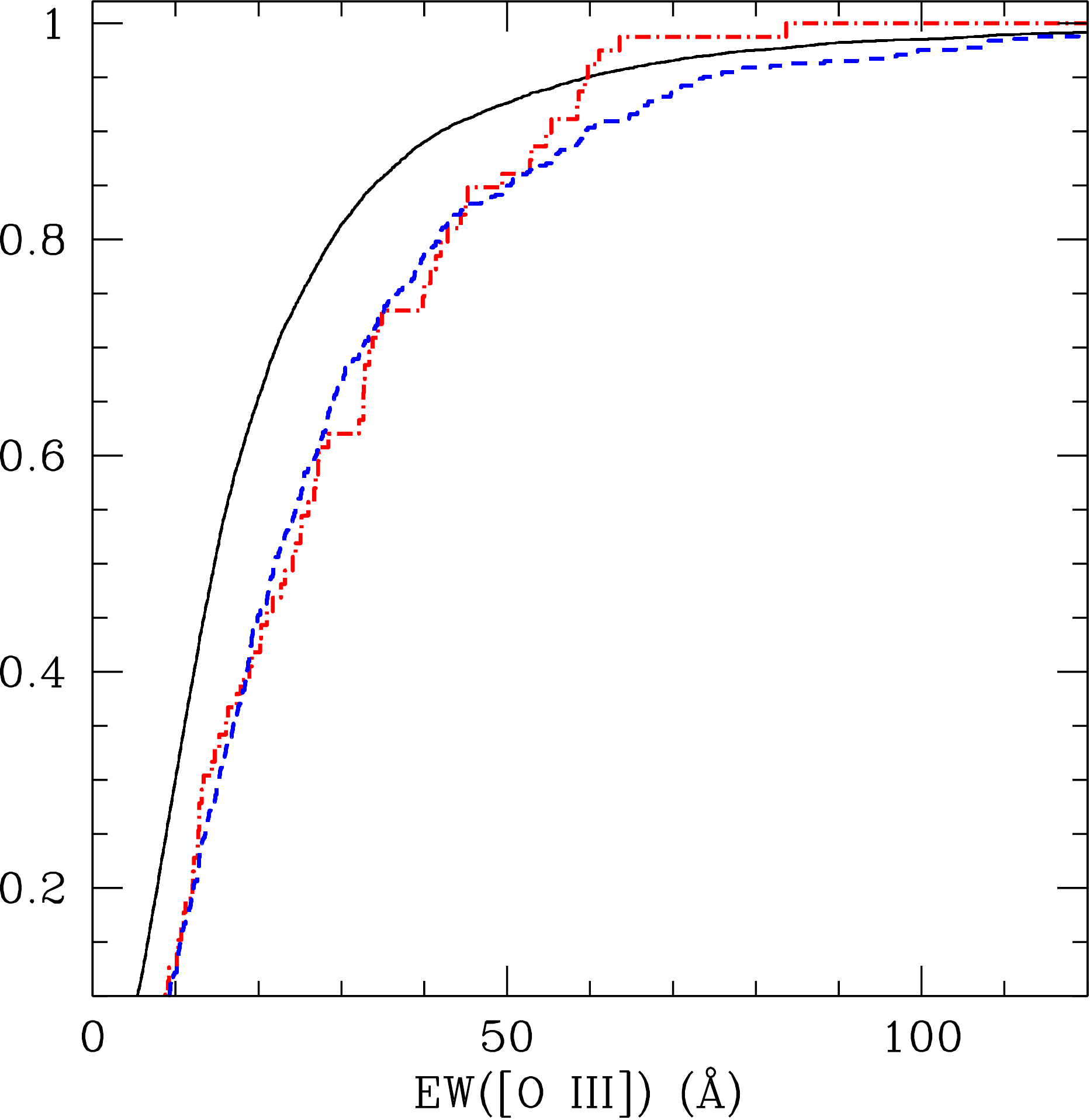}
\caption{Cumulative distributions of \ew\ for the whole sample (black, continuous line), the ``unambiguous'' double-peaked quasars (red, dot-dashed), and the ``possible'' double peaked quasars (blue, dotted line). }
\label{double}
\end{center}
\end{figure}

\subsection{The [OIII] line profiles}

The spectral results on the [OIII] emission line can be interpreted as a simple consequence of the increasing inclination of the sources going from low EWs to high EWs.

In Fig. \ref{oiii_vs_oii} [OIII] and [OII] $3727.092, 3729.875$\AA$\;$ profiles are compared for each of the six stacks. The [OIII] profile shows a prominent blue tail decreasing toward high \ew\, while [OII] holds quite steady in all the bins.

The [OIII] blue component is due to gas in outflow from the NLR towards a direction mostly perpendicular to the plane of the accretion disk. In this scenario, the angle between the outflow direction and the observer line of sight is the same as the disk inclination angle. The velocity component along the line of sight is therefore

\begin{equation}
v_{obs}=v_{outflow} \cos \theta \simeq v_{outflow} \frac{EW[OIII]_{INT}}{EW[OIII]_{OBS}}\;\;.
\end{equation}

The increase in the central velocity shift of the blue [OIII] component with respect to the systemic velocity of the host galaxies ($v_{[OIII]}-v_{[OII]}$) has the same explanation: the shift is more important when the object is face-on because we are observing the outflow exactly along the line of sight (Fig. \ref{voiii-voii} right panel) (the same result was found in \citet{Boroson2011}).

A somewhat more surprising result is the measured blueshift in the [OIII] main component. This finding can be explained as an indirect consequence of orientation effects on the global [OIII] profiles used by \citet{Shen2011} for the estimates of the redshifts. Since we use redshifts from this reference, we are obtaining an inclination dependent systematic shift: more face-on sources have a more prominent blue tail, and so a bluer central $\lambda$ in the global profile. This bias is instead negligible in edge-on objects (Fig. \ref{voiii-voii} left panel).

To give a more quantitative measurement of the [OIII] profile degree of asymmetry we evaluate an asymmetry index similar to that defined in \citet{Heckman1981} and based on differences between Inter-Percentile Velocities (IPV); the asymmetry index is defined as
\begin{equation}
A_{IPV_{05-95}} = \frac{v_{50}-v_{05}}{v_{95}-v_{50}}\;,
\end{equation}
where $v_{05}$, $v_{50}$ and $v_{95}$ are the velocities corresponding to the wavelengths including $5\%$, $50\%$, and $95\%$ of the line total flux.
Through this definition we are able to quantify the asymmetry of a line; with a $A_{IPV_{05-95}}>1$ the line is characterized by a blueward asymmetry, while for $A_{IPV_{05-95}}<1$ the line is more prominent in the red part of its profile.
The asymmetry index for each stack is reported in Tab. \ref{tab5}. $A_{IPV_{05-95}}$ decreases moving towards high \ew\ in agreement with the result in Fig. \ref{OIII_profile} (i.e. a blue tail becoming less prominent at higher \ew.

\begin{figure*}
\centering
\includegraphics[scale=0.42]{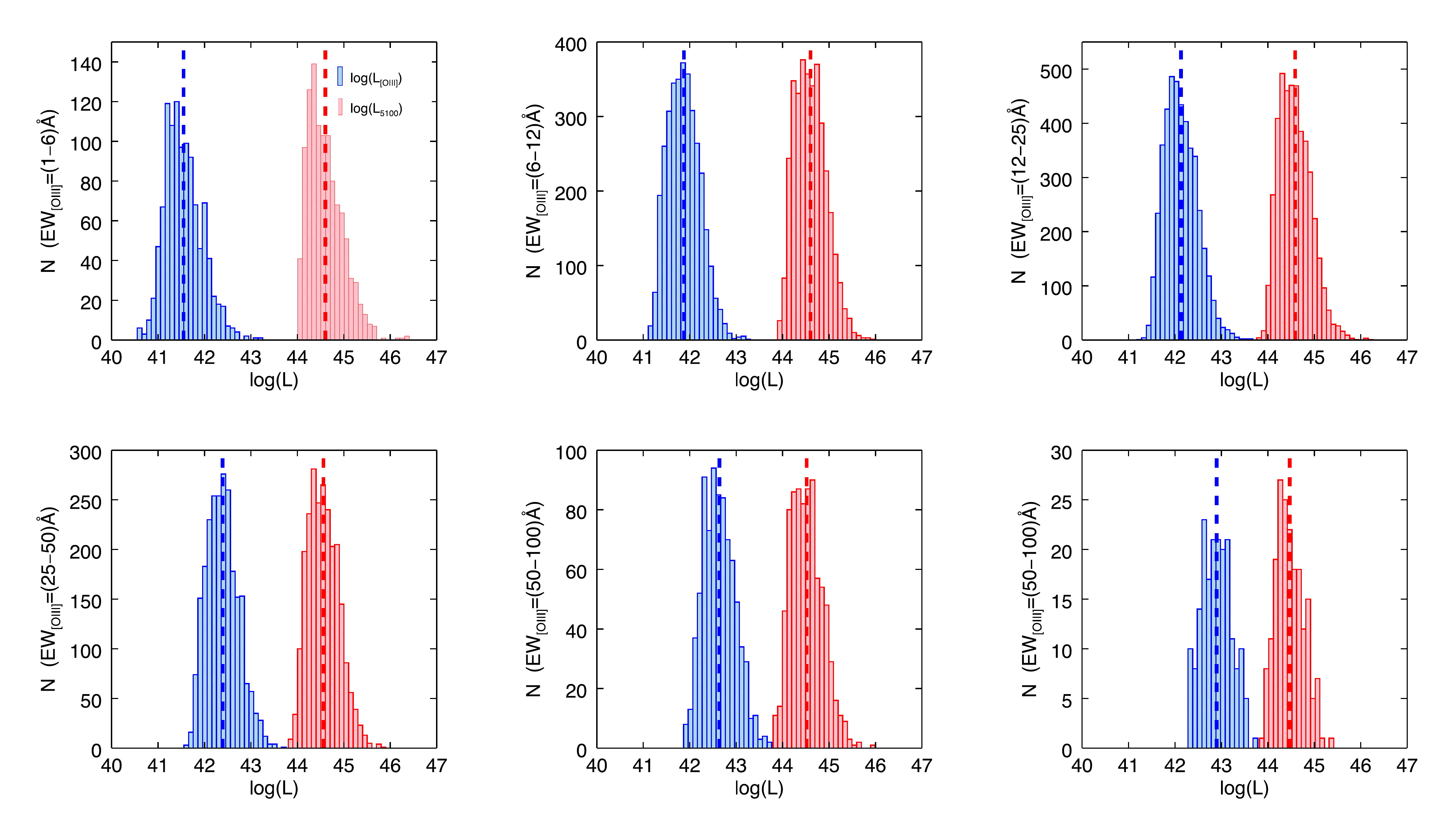}
\caption{$L_{[OIII]}$ and $L_{5100}$ distribution for each stack (the dashed lines represent the mean value for each distribution). The distribution of sources in terms of their continuum at $5100$\AA$\;$ luminosity is stable around $\sim 44.6$, a consequence of the flux limited selection. On the other hand the central value of sources distribution in terms of [OIII] luminosity increases going towards high \ew\ bins. This is due to the flux limit of the sample: when moving towards edge-on positions we are selecting intrinsically more luminous objects.}
\label{hist_Loiii_L5100}
\end{figure*}

\subsection{Broad line profiles}

The number of studies supporting a non-spherical shape of the BLR has constantly grown in recent years.
\citet{Zhu2009} analyzed the BLR profiles in SDSS quasars and suggests that two components with different geometries and physical conditions are needed to reproduce the observed spectra; others studies also suggest the presence of two components, one of them with a spherical geometry while the outer one disk-shaped \citep{Bon2006}. 
In the same BLR disk-shaped scenario it has also been suggested that the kinematics of this inner region, consisting of a combination of rotational and turbulent motions, could affect its geometry, with broader lines emitted from more flattened regions \citep{Kollatschny2011}. Moreover, the proximity of the BLR to the obscuring ``torus'' of the Unified Model \citep{Antonucci93} suggests a smooth connection between the two structures, rather than two completely separated regions, as usually described in the standard unification model. Indeed, the BLR could represent a transition region from the outer accretion disk to the dusty region of the torus \citep{Goad2012}.
Recently \citet{Pancoast2014} used direct modelling techniques on a sample of AGN for which high quality Reverberation Mapping data were available in order to investigate the geometry and the dynamics of the BLR. They found that the geometry of the BLR, as traced by H$\beta$ emission, is consistent with a thick disk.

All these works, despite different aims and explanations, share a common interpretation of the geometry of the BLR.

We claim that the dependence of the broad components on \ew\ is an evidence of the disk-like shape of the BLR: moving from low to high EWs (that is from ``face-on'' to edge-on objects) the component of velocity of the BLR in the direction of the observer grows steadily with the cosine of the inclination angle.

Taking into account this result leads to a number of improvements in our understanding of the AGN inner regions, starting from the determination of the SMBHs virial masses. As long as we consider the BLR as composed by virialized gas, the SMBH mass can be inferred from the BLR lines width according to the relation
\begin{equation}
M_{BH}=f \frac{v_{obs}^{2} R_{BLR}}{G}\;\;,
\end{equation}
where $v$ is the BLR observed line width and $R_{BLR}$ can be obtained from Reverberation Mapping and from the luminosity-$R_{BLR}$ relation for single epoch observations \citep{Kaspi2000, Bentz2013}. An orientation-dependent analysis of emission lines could on one hand improve our knowledge of the morphology of the BLR and so help us in determining more accurately the \emph{virial factor} $f$ \citep{Shen2013}, and on the other hand remove the systematic underestimate of the line widths in non edge-one sources.

\begin{figure*}
\centering
\includegraphics[scale=0.8]{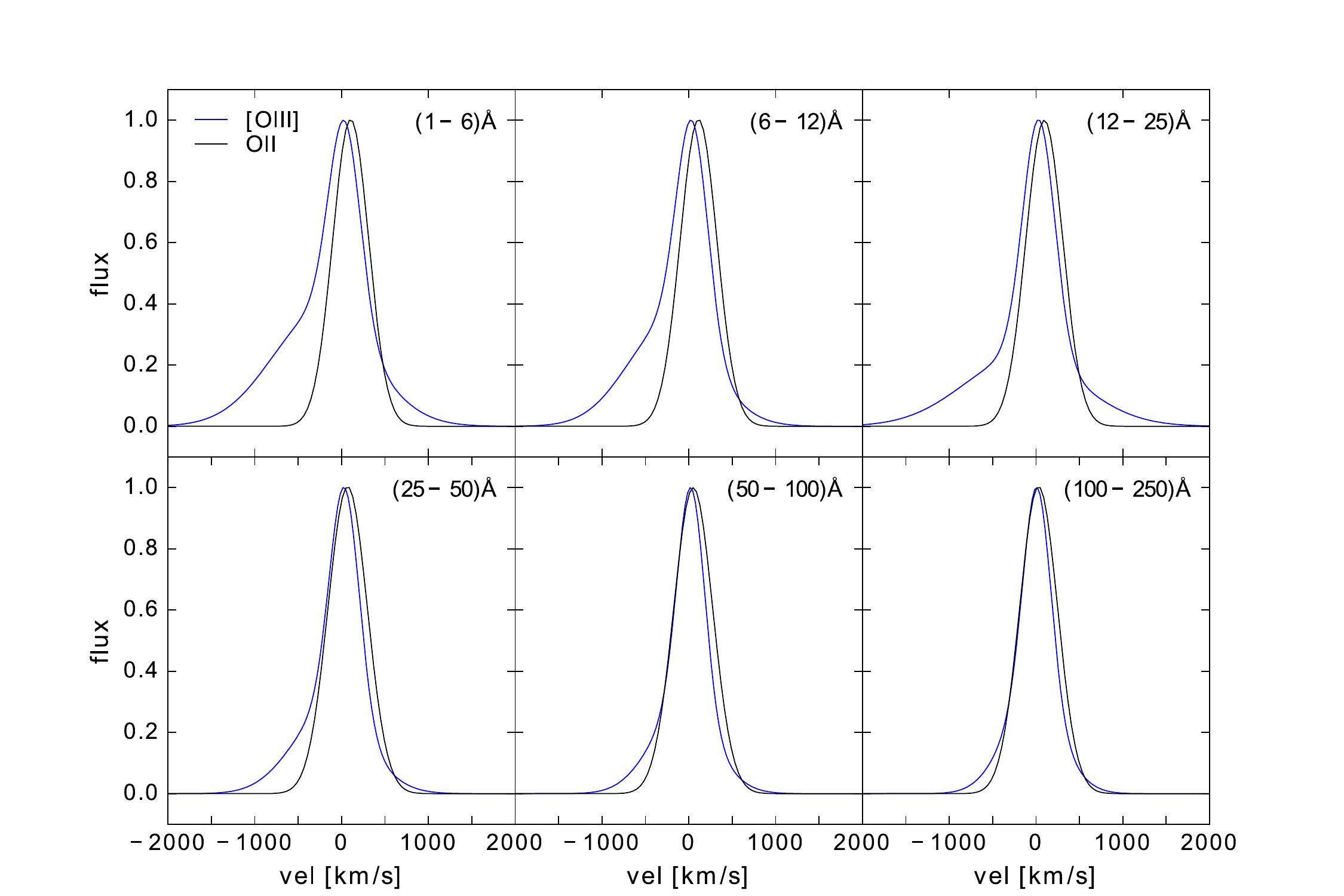}
\caption{[OIII] and [OII] profiles comparison for each stack; the [OIII] blue component decreases moving from low to high \ew\ stacks, i.e. from face-o to edge-on positions. From the relative position of the [OIII] peak with respect to the [OII]  peak we can also estimate the systematic error' in sources redshift: the [OIII] peak is systematically shifted with respect to the [OII] one and the shift is decreasing moving towards higher \ew\ stacks. }
\label{oiii_vs_oii}
\end{figure*}

\subsection{Eigenvector 1}

\begin{figure}
 \centering
   {\includegraphics[scale=0.25]{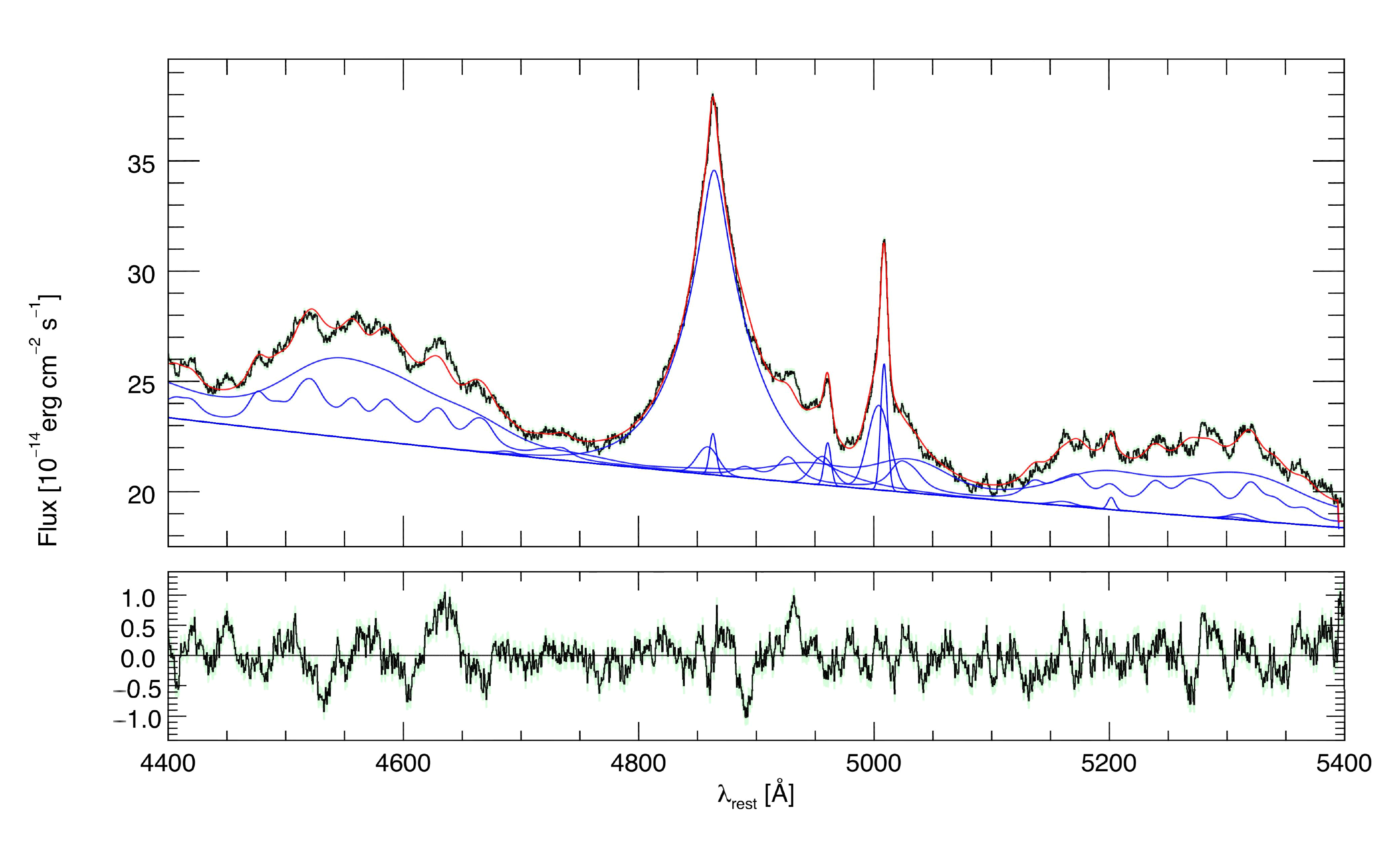}}
   {\includegraphics[scale=0.25]{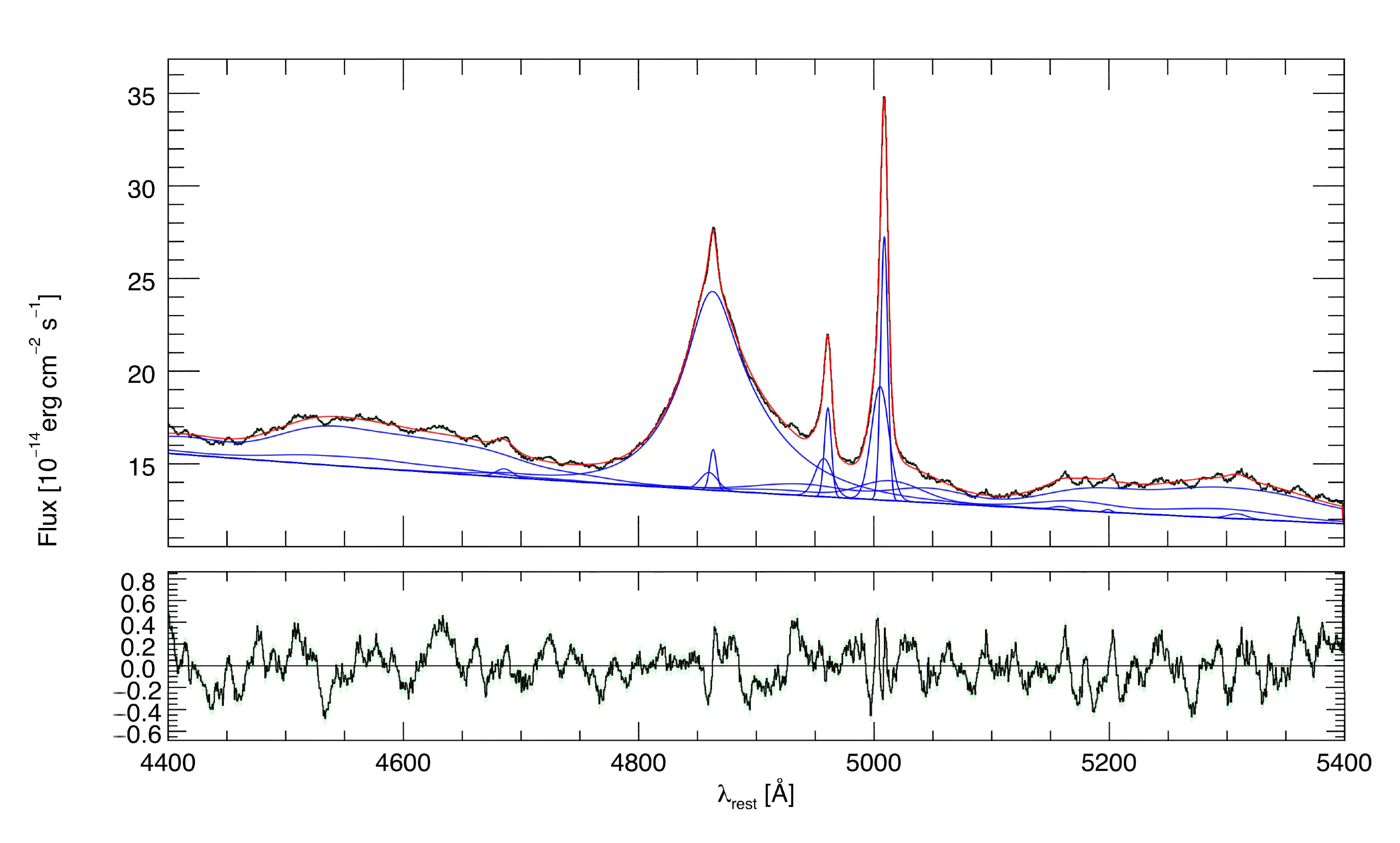}}
   {\includegraphics[scale=0.25]{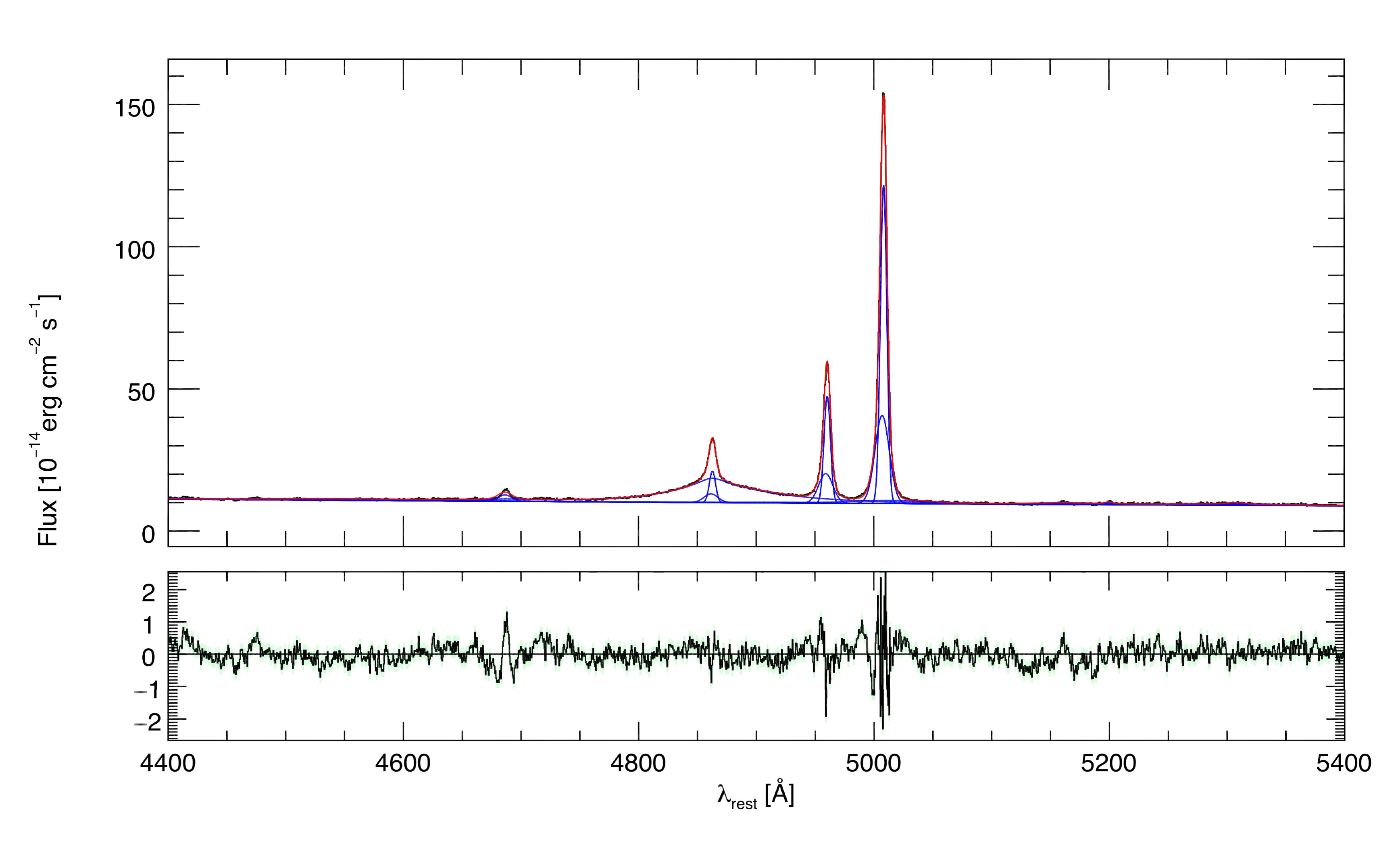}}
 \caption{\protect Fits of the H$\beta$-[OIII] spectral window for the $(1-6)$\AA, $(12-25)$\AA, $(100-250)$\AA$\;$ stacks respectively. The spectrum  is fitted with several functions: for the broad component of permitted lines a double power law convolved with a Gaussian is used. Narrow lines (for both permitted and forbidden lines) are fitted to two Gaussian, the first one accounting for the main component of the line, the other one accounting for the blue tail ascribed to outflowing gas fron the NLR. For the FeII emission several templates are taken into account (see Section \ref{sec:spec_fitting} for details).}
 \label{fig:fits_figures}
\end{figure}

The orientation effects revealed using \ew\ as an inclination indicator provide a possible interpretation of \emph{Eigenvector 1} (EV1) \citep{BorosonGreen1992}, namely the anticorrelation in the intensity of the emissions of FeII and [OIII]. FeII features are BLR lines and so they show the same trend as the other broad lines, i.e. a decreasing intensity and increasing broadening going from face-on (low EWs) to edge-on (high EWs) positions. On the other hand, the [OIII] emission is isotropic. EV1 can then be simply explained in terms of the orientation effects as follows: assuming a disk-like shape for the BLR the intensity of FeII emission lines decreases from face-on to edge-on positions. This effect is clearly present in our stacks: FeII emissions are more prominent in stacks with low \ew\ (Fig. \ref{fig:fits_figures}).
FeII emissions seem to disappear moving towards high \ew, rather than decrease in intensity as the other broad lines do. This fact may suggest that the FeII disk-like structure, besides being flatter than the other broad lines emitting regions, is flatter than the continuum emitting region itself, i.e. the accretion disk. 
Unfortunately the FeII spectrum is characterized by the presence of several, close multiplets and its emissions are by far less intense than those of the other broad lines. This complicates a deeper investigation on this subject. It is possible, in fact, that for the two reasons mentioned above we are simply not able to detect a behaviour of FeII emissions similar to that of the other broad lines.

\section{Summary}

In this paper we analyzed the spectral properties of SDSS quasars as a function of the inclination of the accretion disk with respect to the line of sight. We used the equivalent width of the [OIII] $\lambda5008$\AA~ line as an orientation indicator. Our main results are the following.
\begin{enumerate}
\item[1.] The \ew\ observed distribution is the convolution of an intrinsic log-normal distribution and a power law tail $\propto EW^{-3.5}$, as expected for a randomly distributed population of disks in a flux-limited sample (R11).
\item[2.] The EW(H$\beta$) instead does not show the same tail, suggesting that the BLR could be characterized by a disk-shape geometry.
\item[3.] The R = [OIII]/H$\beta$ distribution resembles the \ew\ observed distribution, with a slightly different value for the power law tail, probably an indication of the deviation of the BLR from a pure disk-like structure.
\item[4.] The [OIII] line has a blue tail, whose intensity and blueshift with respect to the rest frame wavelength both decrease moving from low to high \ew.
\item[5.] All the broad lines behave in the same way: the width of the line increases moving from low to high \ew.
\item[6.] The Eigenvector 1 too has a trend with the \ew\; the FeII emission is strong when \ew\ is low and diminishes gradually while \ew increases.
\item[7.] Double peaked objects are more probable for high \ew\ than for low \ew.
\end{enumerate}
All these findings can be uniquely and satisfactorily explained with the only use of inclination effects; we claim that is very difficult to find an alternative scenario able to account for all these evidences in a similarly simple and straightforward way. 
On the other hand, if the optical/UV of quasars is due to geometrically thin disks, such observational effects are unavoidable.
The scheme we present makes indeed use of only two physical hypothesis: 	
1) the source of the UV-optical continuum is an optically thick and geometrically thin accretion disk, whose luminosity decreases with the cosine of the inclination angle, 2) the [OIII] emission is isotropic and is ascribed to the illumination of the NLR by the same accretion disk.
Such hypothesis naturally arises from our current knowledge on AGN and commonly accerted hypotheses on their nature.

\bibliography{bib_OIIIpap}

\end{document}